# Comparison of Clustering Methods for Extraction of Uncorrelated Sparse Sources from Data Mixtures

Malcolm Woolfson

*Department of Electrical and Electronic Engineering, Faculty of Engineering, University of Nottingham, Nottingham.  NG7 2RD. England*

email: malcolm.woolfson@nottingham.ac.uk

**Abstract**

There is an extensive set of methods to determine sparse sources from mixtures where the mixing coefficients are unknown.  Each method involves plotting $N$ sets of mixed data against each other in $N$-dimensional space.  In the approach adopted in this paper, $N$ dimensional normalised vectors are produced by joining data points that are adjacent in time.  A novel clustering approach is adopted: the two vectors, not necessarily adjacent in time, which are closest to each other are identified and one of these vectors is taken as the principal direction corresponding to one of the sources.  It is shown, using a deflation approach, that it is possible to estimate individual sources to within a multiplicative constant.  This novel method is compared with two related methods and the standard FastICA algorithm.  This new method has comparable performances to three other methods when applied to examples of purely sparse, semi-sparse and non-sparse sources and also when applied to fetal ECG data.

# 1    Introduction

One general problem in signal processing is the extraction of individual source signals $\{s_j[n]\}$ from measurements $\{z_i[n]\}$ that are a linear combination of these sources:

$$z_i[n] = \sum_{j=1}^{N} A_{ij} s_j[n] \quad (1)$$

where $i = 1, 2, \ldots, S$ with $S$ being the number of sets of measurement data, $\{A_{ij}\}$ are the mixing coefficients, and there are $N$ underlying sources. In the case where both the sources and mixing coefficients are unknown, then this problem comes under the heading of Blind Source Separation (BSS). In this paper, we will assume that the number of sources equals the number of measurements, $S = N$.

There are many applications in this area, for example the analysis of EPR data [1], NMR data [2], fetal ECG monitoring [3] and gene mapping [4].

In this paper, we consider the problem where the underlying sources are sparse, for example as shown in Figure 1; in this case, each source is zero for a finite amount of time or, more generally, a particular source is dominant for segments of the data. In many cases, it is not the data itself, but its transform (e.g. Fourier, wavelet) that is sparse.

An example for two sets of sparse data ($N = 2$ in Equation (1)) is shown in Figure 1 below; note that there is a co-incidence of the first peak in each signal.

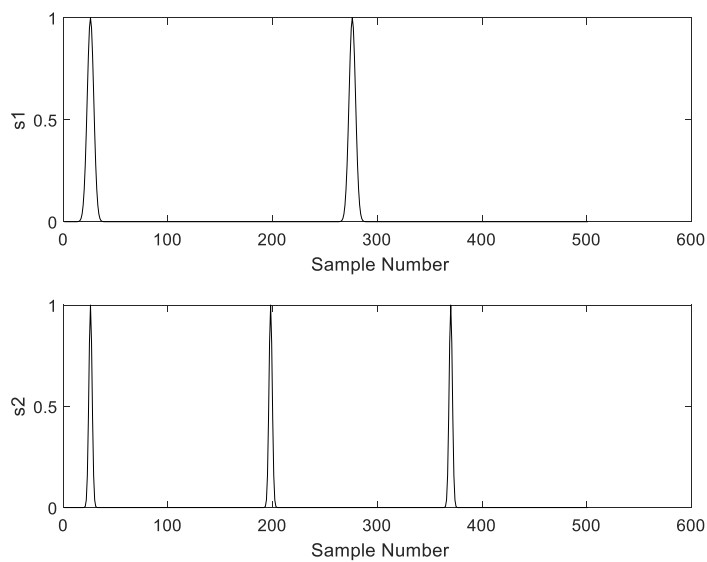

Figure 1 – Two Sparse Sources $s_1$ and $s_2$

Two mixtures of these sources are shown below, where the mixing matrix, which in practical situations is unknown, is given by

$$\mathbf{A} = \begin{pmatrix} 6.5 & 1 \\ 3 & 1 \end{pmatrix}$$

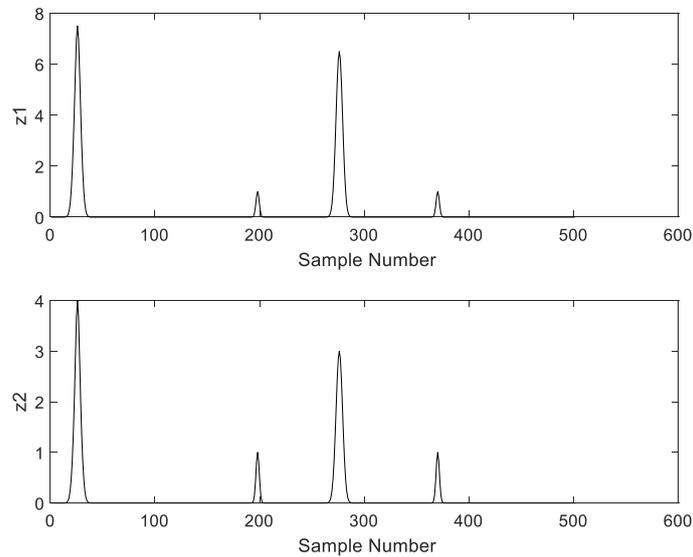

Figure 2 – Mixtures $z_1$ and $z_2$ of sparse sources $s_1$ and $s_2$ in Figure 1

If we plot $z_2[n]$ against $z_1[n]$ for the mixtures shown in Figure 2, then the following plot is obtained:

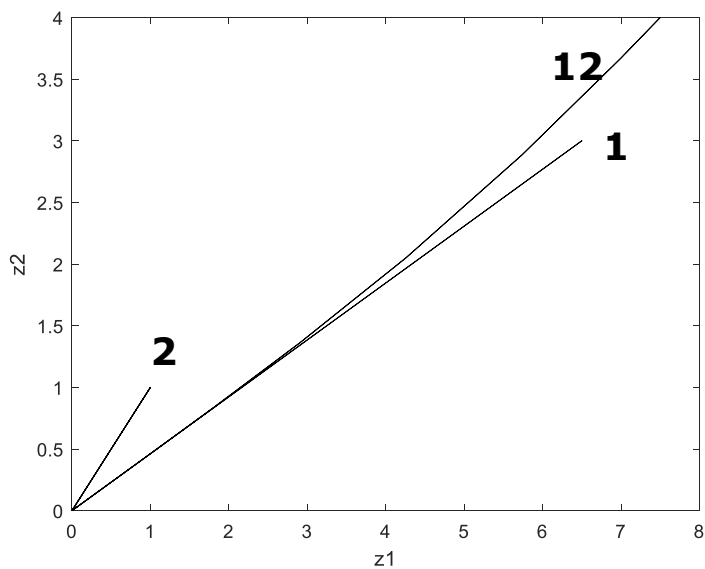

Figure 3 – Plot of $z_2$ against $z_1$ in Figure 2

We will refer to this plot as a *phase space plot*.

During those time segments where $s_2 = 0$ we have from Equation (1):

$$z_1 = A_{11} s_1 \tag{2}$$

$$z_2 = A_{21} s_1 \tag{3}$$

in which case the two measurements are scaled versions of each other.

In this case

$$z_2 = K_1 z_1 \tag{4}$$

where

$$K_1 = \frac{A_{21}}{A_{11}} \tag{5}$$

and the phase space plot will be, locally, a straight line, with the direction corresponding to $s_1$. This is represented by the line marked "1" in Figure 3.

Similarly, for those segments where $s_1 = 0$, then the phase plot will be determined by $s_2$ only and will be a straight line with slope

$$K_2 = \frac{A_{22}}{A_{12}} \tag{6}$$

This is represented by the line marked "2" in Figure 3.

For the example in Figure 2, where there is co-incidence between the peaks of $s_1$ and $s_2$, then the trajectory in the phase plot will not be a straight line; this is marked as "12" in Figure 3.

The aim of this analysis is to find directions in this phase space corresponding to the underlying sources $s_1$ and $s_2$. If, for example, we find the direction at angle θ to the x-axis, corresponding to $s_1$ as indicated in Figure 1 above, then we could rotate axes, so that the *x*-axis is parallel to this direction and $s_1$ is estimated from

$$s_1 = z_1 \cos\theta + z_2 \sin\theta \tag{7}$$

If $s_1$ and $s_2$ are uncorrelated and $z_1$, $z_2$ have been normalised to unit variance, then the direction corresponding to $s_2$ in the phase plot is perpendicular to $s_1$. By transforming the data to these new axes ($s_1, s_2$) one can then extract the sources to within a multiplicative constant.

There are many methods addressing this problem of finding the dominant directions in phase space. Statistical methods have been used to find the dominant directions based on, for example, Principal Components Analysis (PCA) [5], Independent Component Analysis (ICA) [6,7] and Blind Source Separation - Maximum Likelihood (BSS-ML) [8]. The above-mentioned methods are appropriate for general sources which have non-zero values for all times, and which are not necessarily sparse. The general problem of extracting sparse sources has been looked at by various workers, for example in References [9-27]. The main approach of Sparse Component Analysis (SCA) is to identify segments of the data where one source exists on its own, or at the very least dominates the other sources, and use this information to estimate the direction in phase space corresponding to that source. Another application of SCA is to the underdetermined case where the number of data inputs is less than the number of sources. Any sparsity in the underlying sources can be used to estimate the mixing matrix and hence the sources.

Examples of clustering methods used to find these principal directions are K-means [26], clustering of angles [27] and clustering of normals of hyperplanes [20]; references to other clustering methods can be found in [20].

In [17], the Minimum Heading Change (MHC) method is developed. In this and subsequent papers [18,25], the phase plots are thought of as being like the trajectories of aircraft on a radar screen. Looking, for simplicity, at the two-dimensional case, we can consider the point ($z_1$,$z_2$) in phase space as corresponding to the position co-ordinates of an aircraft on the radar screen. Adjacent points can be joined up to produce "velocity" vectors as would be seen on a radar screen and the normalised velocity vectors could then be considered as "headings". This is an unconventional way of thinking of the data, but it can be useful when constructing algorithms to identify segments of the trajectories corresponding to one signal source only. In References [17] and [18], the dominant heading directions are determined by the magnitude of the difference between two adjacent normalised headings in time being less than a certain threshold. In this case, this segment would consist of one source only, corresponding to the segments marked "1" and "2" in Figure 3. This is a straightforward method to implement, but this can be sensitive to noise especially if one of the velocity vectors involved has small magnitude. In addition, this method only considers the similarity between headings that are adjacent in time; if a particular source is sparse at non-adjacent data points, then pairs of headings will not be grouped together, and some useful information will be lost.

An alternative approach to the MHC is described in [25], where a "Global" method is employed, where heading components are grouped together across the whole data interval; there is no requirement for headings within a group to be adjacent in time. Heading clusters are formed such that within each cluster, differences in magnitude between clustered heading components are within a threshold. Within each cluster, maximum likelihood averaging of headings (considering the magnitude of the corresponding velocity vectors) is carried out to estimate the dominant directions in phase space; in principle, because of this averaging, this method should be more robust to noise than the MHC method if there are more than two headings within a cluster. However, the performance of this method depends on the choice of threshold for clustering: if the threshold is too large, heading vectors will be included which correspond to more than one source; if the threshold too small then important headings will be missed and not included in the averaging. We will call this method Global 1.

In this paper, we investigate another method of clustering where we choose the pair of headings across all the data, with the minimum magnitude of the difference between them, to detect sparse sources. No thresholding of heading differences is required here, which could make the method more robust. We will call this method Global 2.

The aim of this paper is to compare the performances of the MHC, Global 1 and Global 2 methods when applied to a variety of simulated and experimental data to determine where each method works best. Robustness to noise will be looked at. A comparison with the standard Fast ICA method [6,7] will also be made.

## 2  Estimation of Sources

Two methods of estimating the sources from the headings or principal directions in phase space are described below. In the first approach, the Direct Method, all the principal directions in phase space are estimated, and the estimated sources can be found by a direct matrix inversion. In the second approach, the deflation method, the principal directions are estimated one-by-one, at each stage estimating the corresponding source and the subtracting off the contribution of that source from the data.

### 2.1  Direct Method

Once, the dominant directions are found in phase-space then one can determine the sources to within a scaling constant by inverting the mixing matrix.

From Equation (1) where $s_2 = 0$:

$$z_1^{(1)} = A_{11}s_1 \tag{8}$$

$$z_1^{(2)} = A_{21}s_1 \tag{9}$$

The position in phase space corresponding to Source 1 is given by the vector

$$\boldsymbol{r_1} = (A_{11}s_1, A_{21}s_1) \tag{10}$$

which can be normalised to remove the dependence on $s_1$. Hence the determination of this direction gives information about the first column of the mixing matrix.

Similarly, the position in phase space corresponding to Source 2 is given by the vector

$$\boldsymbol{r_2} = (A_{12}s_2, A_{22}s_2) \tag{11}$$

which can be normalised to remove the dependence on $s_2$ and hence information on the second column of the mixing matrix can be obtained.

It can be shown that the relation between the mixed data (measurements) and the underlying sources is given by a relation of the form

$$\boldsymbol{z} = \boldsymbol{A'}\boldsymbol{s'} \tag{12}$$

where $\boldsymbol{A'}$ and $\boldsymbol{s'}$ are modified versions of the mixing matrix and source vector taking into account the normalisation of $\boldsymbol{r_1}$ and $\boldsymbol{r_2}$.

The sources can then be estimated, to within an unknown multiplicative constant, from:

$$\boldsymbol{s'} = \boldsymbol{A'}^{-1}\boldsymbol{z} \tag{13}$$

Thus, there are two stages in the estimation of sources: (i) estimation of the mixing matrix $\boldsymbol{A'}$ in (12) and (ii) estimation of the sources from (13). This approach is used in several methods of SCA.

This method has a good performance if each source is non-zero for a segment of time. There is no requirement, when using this approach, for the sources to be uncorrelated. The disadvantage of this approach is that it would not work for sources that are hidden by other sources, or if some sources are weaker than others and are difficult to detect. This approach requires a reliable method to determine directions in phase space for <u>all</u> sources. If any sources are undetected or if false sources are detected, then this will affect the estimation of all sources.

**2.2 The Deflationary Approach to Signal Estimation**

An alternative to the direct method described in Section 2.1 is the deflationary approach, where one determines directions in phase space corresponding to each source one at a time; this is the approach that has been used in, for example, [17],[18] and [25]. and is also an option in FastICA. As each source is detected and estimated, it is subtracted off the data and the method is iterated to detect and estimate further sources. This is the process adopted in this paper. The advantage of the deflationary approach is that one can deal with problem of hidden sources and small amplitude sources as more significant sources are detected first. The disadvantage of this approach is that the application of repeated subtractions makes the method more sensitive to noise. In addition, we will see that for this method to extract the sources without error, it is a requirement that the sources are uncorrelated, a condition not required by the direct method described in Section 2.1.

The first step is to "whiten" the components, that is form a set of components that are orthonormal. This procedure is useful in that uncorrelated sources are represented by orthogonal directions in phase space, and we will see later that this can lead to all sources being extracted to within a multiplicative constant. In the methods described in this paper, the method of whitening that is chosen is not important, but, for simplicity, we choose the Gram-Schmidt procedure. Each Gram-Schmidt component is normalised to unity amplitude. The deflationary approach does not work well for highly correlated sources, because directions of correlated sources in phase space are not orthogonal after whitening of the data; this means that the estimate of one source will pick up contributions from other sources.

When implementing the PCA and BSS methods one common first step is to subtract off the dc component from the original data set. We shall see later that this can be undesirable in some cases as this may introduce unintended correlations into the assumed underlying sources.

In this section, the deflation method is described for the case where there are $N$ underlying sources $\{s_i(t)\}$, i=1,2,…,$N$, under the following assumptions:

(1)     The number of mixed data signals equals the number of sources

(2)     For each source there is a specific time slot such that all the other sources are zero.

In this analysis, lower case symbols refer to variables that are a function of time, whilst upper case symbols refer to time-independent variables. For notational convenience, the time-dependencies are suppressed. Note that this notation means that some vectors will be represented by upper case variables, but the vector nature of such variables will be emphasised.

The mixed signals, which form the measurements, are assumed to be given by Equation (1):

$$z_i = \sum_{j=1}^{N} A_{ij} s_j \tag{14}$$

where $\{A_{ij}\}$ ($i=1,\ldots,N; j = 1,\ldots,N$) are mixing coefficients.

Following the procedure adopted in conventional BSS methods, the components are whitened to obtain the following transformed measurements:

$$e_i = \sum_{j=1}^{N} B_{ij} s_j \tag{15}$$

where ($i=1,2,\ldots N$), and $\{B_{ij}\}$ are constants depending on the whitening procedure used – in the approach described in this paper, we use the Gram-Schmidt method with normalisation of the components.

Define the time varying vector of transformed measurements (15) in $N$ dimensional phase space as

$$\boldsymbol{e} = (e_1, e_2, \ldots, e_N) \tag{16}$$

The assumption of sparsity of all sources is made, which is that, during certain segments of time, for each component $p$ $s_p \neq 0$ and all the other components are zero: $s_i = 0$ for all $i \neq p$. In these situations, the phase space plot corresponding to these time slots are straight lines.

When $s_i = 0$ for all $i \neq p$, from Equation (15) the whitened components become $\{e_i^{(p)}\}$ where

$$e_i^{(p)} = B_{ip} s_p \quad (i = 1,\ldots,N) \tag{17}$$

In this case, the phase space plot is given by the following vector

$$\boldsymbol{s_p} = (e_1^{(p)}, e_2^{(p)}, \ldots, e_N^{(p)}) = s_p(B_{1p}, B_{2p}, \ldots, B_{Np}) = s_p \boldsymbol{R_p} \tag{18}$$

where

$$\mathbf{R_p} = (B_{1p}, B_{2p}, \ldots, B_{Np}) \tag{19}$$

is an (1 x $N$) vector that represents a constant direction in phase space during the time that $s_i = 0$ for all $i \neq p$ ; as an example, for the case of mixtures of two sources, this is represented by segments 1 and 2 in Figure 3, where $s_2 = 0$ and $s_1 = 0$ respectively. In the regions where $s_1$ and $s_2$ are both non-zero, then the phase plot looks like segment 12 in Figure 3 i.e., not a straight line.

Using Equation (15) we may write for the general whitened vector **e**:

$$\mathbf{e} = (e_1, e_2, \ldots, e_N) = (\sum_{j=1}^{N} B_{1j}s_j, \sum_{j=1}^{N} B_{2j}s_j, \ldots, \sum_{j=1}^{N} B_{Nj}s_j) \tag{20}$$

We may expand this expression as:

$$\mathbf{e} = (B_{11}s_1, B_{21}s_1, \ldots, B_{N1}s_1) + (B_{12}s_2, B_{22}s_2, \ldots, B_{N2}s_2) + \ldots + (B_{1N}s_N, B_{2N}s_N, \ldots, B_{NN}s_N) \tag{21}$$

which can be further rewritten as:

$$\mathbf{e} = (B_{11}, B_{21}, \ldots, B_{N1})s_1 + (B_{12}, B_{22}, \ldots, B_{N2})s_2 + \ldots + (B_{1N}, B_{2N}, \ldots, B_{NN})s_N \tag{22}$$

Using Equations (19) and (22) **e** can be written in the form:

$$\mathbf{e} = \sum_{i=1}^{N} s_i \mathbf{R}_i \tag{23}$$

The main purpose of the algorithm is to identify the direction in phase space, specified by the unit vector $\widehat{\mathbf{R}}_P$, corresponding to each source $s_p$ ($p= 1,2,\ldots N$) in the phase plot. These vectors can be found when all the other sources are zero for a short segment of data. In Section (3), we will describe three methods, MHC, Global 1 and Global 2, to estimate these vectors. For the moment, we will assume that the correct principal direction is detected by one of these methods.

Suppose that the direction corresponding to source $p$, $\mathbf{R}_p$, is detected.

Let the unit vector in this direction be $\widehat{\mathbf{R}}_p$.

From Equation (19):

$$\widehat{\boldsymbol{R}}_P = \frac{(B_{1p}, B_{2p}, \ldots, B_{Np})}{\sqrt{B_{1p}^2 + B_{2p}^2 + \ldots + B_{Np}^2}} \tag{24}$$

An estimate of this source signal, $\tilde{s}_p$, is obtained from the phase plot by finding the component of the phase plot data in this direction:

$$\tilde{s}_p = \widehat{\boldsymbol{R}}_P . \boldsymbol{e} \tag{25}$$

Substituting for **e** from Equation (23):

$$\tilde{s}_p = \widehat{\boldsymbol{R}}_p . \sum_{i=1}^{N} \boldsymbol{R}_i s_i \tag{26}$$

Writing

$$\boldsymbol{R}_i = R_i \widehat{\boldsymbol{R}}_i \tag{27}$$

Equation (26) can be rewritten as

$$\tilde{s}_p = \widehat{\boldsymbol{R}}_p . \sum_{i=1}^{N} \widehat{\boldsymbol{R}}_i R_i s_i \tag{28}$$

After further rearranging Equation (28) can be written as:

$$\tilde{s}_p = \sum_{i=1}^{N} s_i R_i (\widehat{\boldsymbol{R}}_p . \widehat{\boldsymbol{R}}_i) \tag{29}$$

In the special case where all the source vectors are uncorrelated, and whitening has taken place, $\widehat{\boldsymbol{R}}_p . \widehat{\boldsymbol{R}}_i = 0$ for all $i \neq p$ in Equation (29) and

$$\tilde{s}_p = s_p R_p \tag{30}$$

and source $s_p$ is extracted to within a scaling constant $R_p$. In the general case where $s_p$ is correlated with some other sources then $\tilde{s}_p$ will depend on both $s_p$ and these other sources.

In the phase space plot, from Equations (27) and (30), the estimate of source $s_p$, $\tilde{s}_p$, is represented by a vector

$$\tilde{\boldsymbol{s}}_p = \tilde{s}_p \widehat{\boldsymbol{R}}_P \tag{31}$$

Substituting for $\tilde{s}_p$ from Equation (29) into (31):

$$\tilde{s}_p = \widehat{R}_P \sum_{i=1}^{N} s_i R_i (\widehat{R}_p \cdot \widehat{R}_i) \tag{32}$$

Separating out the term $i = p$ in the summation:

$$\tilde{s}_p = s_p R_p \widehat{R}_P + \widehat{R}_P \sum_{i=1(i \neq p)}^{N} s_i R_i (\widehat{R}_p \cdot \widehat{R}_i) \tag{33}$$

We now subtract this vector from the original whitened vector **e:**

$$e' = e - \tilde{s}_p \tag{34}$$

as a first step to find the other sources.

Using Equations (23), (27) and (33):

$$e' = \sum_{i=1}^{N} s_i R_i \widehat{R}_i - s_p R_p \widehat{R}_p - \widehat{R}_p \sum_{i=1(i \neq p)}^{N} s_i R_i (\widehat{R}_p \cdot \widehat{R}_i) \tag{35}$$

Combining the first two terms into one:

$$e' = \sum_{i=1(i \neq p)}^{N} s_i R_i \widehat{R}_i - \widehat{R}_p \sum_{i=1(i \neq p)}^{N} s_i R_i (\widehat{R}_p \cdot \widehat{R}_i) \tag{36}$$

which can be further rewritten as:

$$e' = \sum_{i=1(i \neq p)}^{N} s_i R_i [\widehat{R}_i - \widehat{R}_p (\widehat{R}_p \cdot \widehat{R}_i)] \tag{37}$$

Equation (37) can be written in the more compact form:

$$e' = \sum_{i=1(i \neq p)}^{N} s_i R'_i \tag{38}$$

where

$$R'_i = R_i [\widehat{R}_i - \widehat{R}_p (\widehat{R}_p \cdot \widehat{R}_i)] \tag{39}$$

In this case the time varying vector $e'$ in Equation (38) does not depend on the source $s_p$ which has been subtracted out of the calculation; this vector depends on the other $N$-1 sources that have not yet been estimated. The above procedure is iterated until there are no components left. Our final estimates for the sources are $\{\tilde{s}_p\}$ in Equation (30) where $p = 1,2,\ldots,N$. It should be noted that the final source is estimated exactly to within a multiplicative constant, as all other sources have (ideally) been subtracted out.

Comments:

(1)  The above analysis assumes that there is an interval of time when each source is non-zero with the other sources being zero. At the last iteration we are left with one of the source vectors with no contributions from other sources.

(2)  In the special case where all the sources are uncorrelated and the data has been pre-whitened $\widehat{\boldsymbol{R}}_p \cdot \widehat{\boldsymbol{R}}_i = 0$ for $(i \neq p)$ then, from Equation (33), at each iteration a specific source signal is extracted uncontaminated by other source signals and the vector representing each source is given by:

$$\tilde{\boldsymbol{s}}_p = s_p R_p \widehat{\boldsymbol{R}}_P \tag{40}$$

in which case the estimate of the source itself is

$$\widetilde{s_p} = \tilde{\boldsymbol{s}}_p \cdot \widehat{\boldsymbol{R}}_P = s_p R_p \tag{41}$$

## 2.3 Preprocessing

### 2.3.1 Calculating "Velocity Vectors"

Now we could look for principal directions in phase space just using the data $\{e_i[n]\}$. However, dc shifts in the data would mean that a principal source will be represented by a vector that changes its direction in time. To circumvent this problem, for each component $i$ we calculate "velocity" vectors from adjacent data points:

$$\boldsymbol{v}[n] = (v_1[n], v_2[n], \ldots, v_N[n]) \tag{42}$$

where

$$v_i[n] = e_i[n] - e_i[n-1] \tag{43}$$

This differencing, which occurs after whitening of the data, subtracts out any dc components without introducing correlations into the data which could occur if we subtracted the average out of the original data prior to whitening. The principal directions in the data are found from the points in phase space defined by the set of velocity vectors

$$\{\boldsymbol{v}[1], \boldsymbol{v}[2], \ldots, \boldsymbol{v}[M]\} \tag{44}$$

where $M$ is the number of data points.

### 2.3.2 Thresholding of Velocity Vectors

We now discuss how we deal with the problems of noise affecting the input data.

Let the $N$ components of the velocity vector at time point $n$ be written as

$$\boldsymbol{v}[n] = (v_1[n], v_2[n], \ldots, v_N[n]) \tag{45}$$

Noise will affect small magnitude velocity vectors more than the larger ones, so we threshold the data so that we accept velocity vectors that have a magnitude larger than a certain threshold.

In order to avoid effects of spurious noise, a velocity vector is accepted at sample point $n$ if

$$V_{max}[n] \geq v^{th} \cdot v_{max} \tag{46}$$

where

$$V_{max}[n] = \max\{|v_1[n]|, |v_2[n]|, \ldots, |v_N[n]|\} \tag{47}$$

$0 < v^{th} < 1$ is a chosen threshold

and

$$v_{max} = \{|\boldsymbol{v}[1]|, |\boldsymbol{v}[2]|, \ldots, |\boldsymbol{v}[M]|\}_{max} \tag{48}$$

is the maximum value of the magnitude of the velocity vector over all $M$ sample points. This threshold is used in both the MHC and Global methods to be discussed next.

### 3. Methods to Detect and Estimate Headings

For all the phase space methods described in this paper, the first three steps are common:

Step 1:   Input whitened data $\{e_i[n]\}, (i = 1, \ldots, N)$

Step 2:   For each component $i$ calculate velocity vectors from adjacent data points (43):

$\mathbf{v}[n] = \{v_i[n]\} = \{e_i[n] - e_i[n\text{-}1]\}$

To illustrate the various methods used in this work, we will use the following simplified example where there are 10 headings to sort into clusters and the number of sources, $N = 2$, with the velocity components $(v_1[n], v_2[n])$ given in Table 1.

| Heading Number, n | $v_1[n]$ | $v_2[n]$ |
|---|---|---|
| 1 | 1 | 2 |
| 2 | -2 | 3 |
| 3 | 1 | 2 |
| 4 | -2 | -4 |
| 5 | 5 | 3 |
| 6 | -1 | -2 |
| 7 | -4 | 6 |
| 8 | 5 | 5 |
| 9 | -4 | 6 |
| 10 | 5 | 10 |

*Table 1 – Velocity Components*

Step 3 : We now calculate the heading vector according to

$$\hat{\mathbf{r}}[n] = \frac{\mathbf{v}[n]}{|\mathbf{v}[n]|} \tag{49}$$

| Normalised Heading Number, n | $\hat{r}_1[n]$ | $\hat{r}_2[n]$ |
|---|---|---|
| 1 | 0.4472 | 0.8944 |
| 2 | -0.5547 | 0.8321 |
| 3 | 0.4472 | 0.8944 |
| 4 | -0.4472 | -0.8944 |
| 5 | 0.8575 | 0.5145 |
| 6 | -0.4472 | -0.8944 |
| 7 | -0.5547 | 0.8321 |
| 8 | 0.7071 | 0.7071 |
| 9 | -0.5547 | 0.8321 |
| 10 | 0.4472 | 0.8944 |

*Table 2 – Heading Vectors*

It is now clear that there are two heading clusters: (1,3,4,6,10) and (2,7,9) and we want to detect one of these to estimate the corresponding source and to subtract from the data. We now describe three methods to do this.

### 3.1  Minimum Heading Change (MHC) Method [17]

#### 3.1.1  Principle of Method

Determine the normalised headings in the phase space plot

Starting off at $n = 1$ and going down the list, the heading difference is calculated according to

$$\boldsymbol{d}_+[n] = \hat{\boldsymbol{r}}[n] + \hat{\boldsymbol{r}}[n-1] \tag{50}$$

and

$$\boldsymbol{d}_-[n] = \hat{\boldsymbol{r}}[n] - \hat{\boldsymbol{r}}[n-1] \tag{51}$$

Note that we need to calculate $\mathbf{d}_+[n]$ to take into account sign inversion of headings.

At each time point n, work out $\varepsilon[n]$ as follows:

$$\varepsilon[n] = \min[|\boldsymbol{d}_+[n]|, |\boldsymbol{d}_-[n]|] \tag{52}$$

In this case $\varepsilon[n]$ is the difference in heading magnitudes between adjacent headings.

Using the example in Table 2, the following values of $\varepsilon[n]$ are obtained:

| $n$ | $\varepsilon[n]$ |
|---|---|
| 1 | x |
| 2 | 1.0038 |
| 3 | 1.0038 |
| 4 | 0 |
| 5 | 0.5592 |
| 6 | 0.5592 |
| 7 | 1.0038 |
| 8 | 1.268 |
| 9 | 1.268 |
| 10 | 1.0038 |

*Table 3 – Differences between headings*

Note that $\varepsilon[1]$ is undefined as indicated by an "x" in the above Table.

$\varepsilon[n]$ is a minimum when $n = 4$, so heading 4 is chosen, which we have already noted is a member of one of the clusters.

The advantage of this method is that it is relatively straightforward to implement; all one does is to perform a simple addition and subtraction of two adjacent normalised vectors.

A disadvantage of this method is that it relies on the sources being sparse enough so that one can find headings in a cluster that are adjacent in time. In the above example, although there are two clusters with 5 and 3 headings as members, only two adjacent headings belong to a cluster. In the worst case, no headings in a cluster may be adjacent, in which case the wrong vector may be chosen corresponding to a source.

### 3.1.2 Simulations

The MHC has been applied to the data in Figure 2, where the threshold $v^{th} = 0.1$ has been chosen in Equation (46).

Figure 4 shows vertical lines indicating the sample points where the principal headings have been detected at the first and second iteration; the method correctly detects Source 2 at the first iteration (sample point 198) and Source 1 (sample point 276) at the second iteration.

**First Iteration**

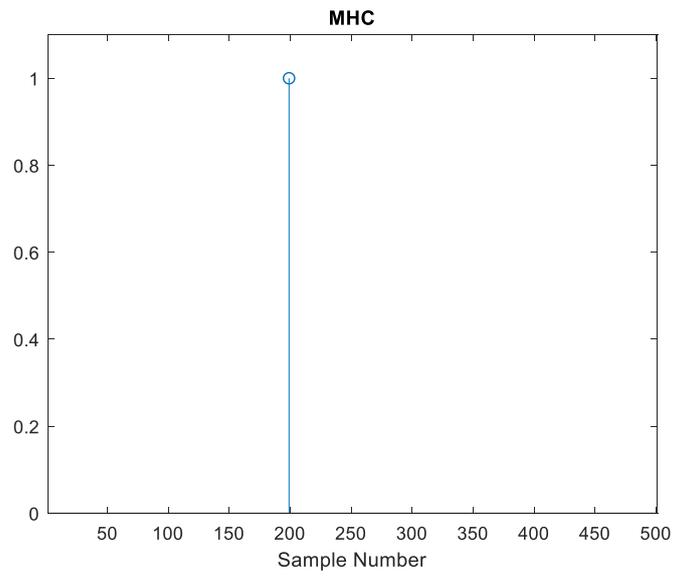

**Second Iteration**

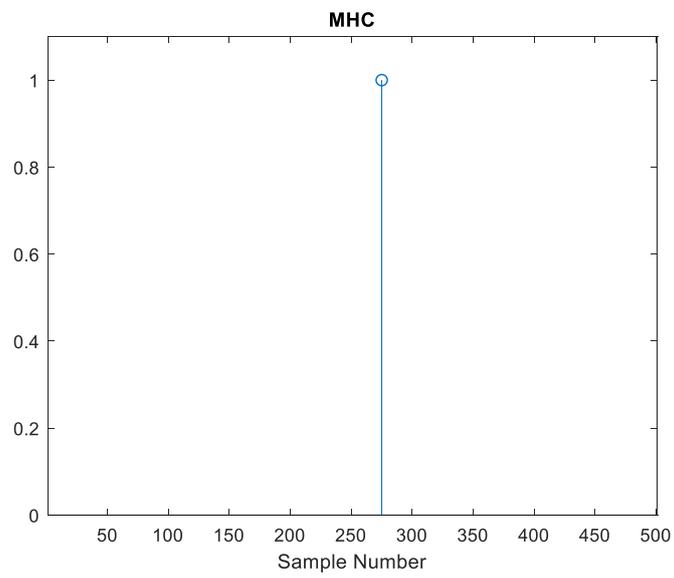

Figure 4 – Samples where Sources are Detected using MHC

The estimated sources are shown in Figure 5:

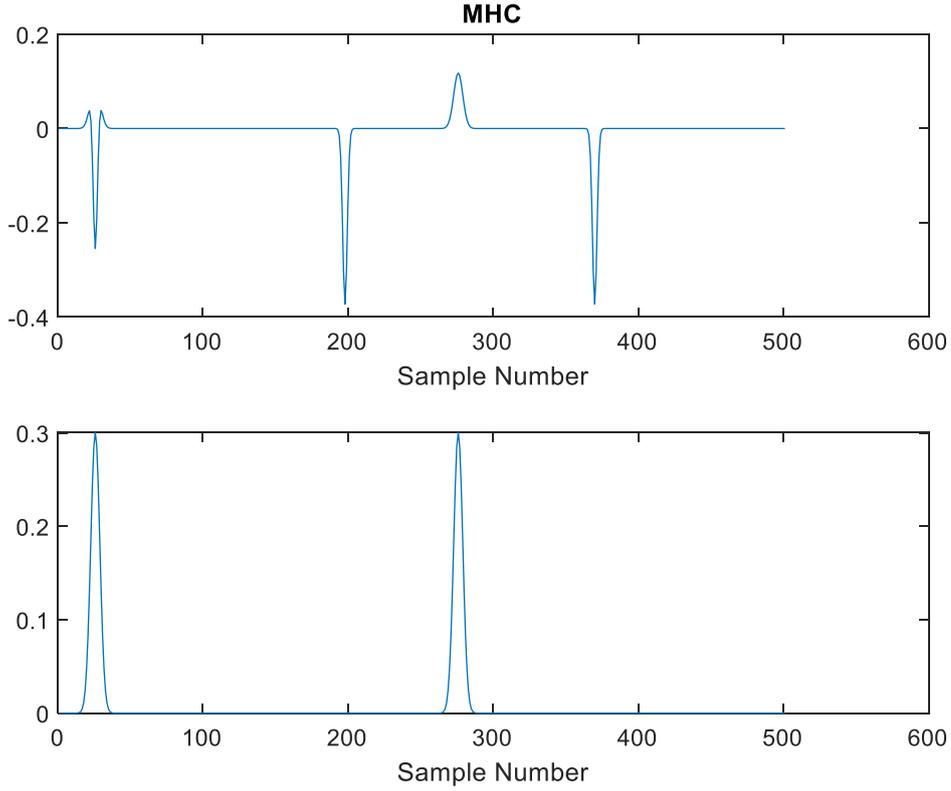

Figure 5 – Estimated Sources. Top: first iteration, Bottom: second iteration.

Note that the estimate of Source 2 (in top half of Figure 5) is contaminated by Source 1; this is expected because, from Figure 1, there is coincidence between the first peaks of both sources and therefore the underlying sources are correlated. This means that the vectors in phase space representing each source are not orthogonal.

From Equation (33), with $N = 2$ and $p = 2$:

$$\tilde{s}_2 = s_2 R_2 \widehat{R}_2 + \widehat{R}_2 s_1 R_1 (\widehat{R}_2 . \widehat{R}_1) \tag{53}$$

Now for correlated sources, the scalar product $\widehat{R}_2 . \widehat{R}_1 \neq 0$ and so $s_1$ contributes to the estimate of $s_2$ as observed in the top graph of Figure 5. It can be observed from the bottom graph in Figure 5 that Source 1, which is detected in the second iteration, is estimated to within a multiplicative constant; this was predicted in Comment (1) in Section (2.2) above.

### 3.2 Clustering Methods

A potential improvement to the MHC method described in Section 3.1.1, is to cluster headings in the dataset that are not necessarily adjacent in time. This would occur if the same source were sparse for different segments of data.

The presence of noise in the input data will cause a problem for the MHC in detecting these directions. Only one heading value is chosen in this direction which may deviate significantly from the actual heading. The question to be asked is whether one can obtain more robustness to noise if one clusters heading vectors over the whole phase plot corresponding to each dominant source direction, similar to the clustering of points used in [16]. these heading vectors need not be adjacent in time. One could then perform a weighted average over the heading vectors to obtain a smoother estimate of the underlying sources. One such method is described in [25], which we will refer to as Global 1 and, for completeness, this is described below. Later, we derive a suboptimal method, called Global 2, which requires less input parameters than Global 1.

### 3.2.1 Global 1 [25]

*Detection of Clusters*

A basic method is to directly compare heading components at different time points and to calculate the magnitude of the difference between pairs of vectors. The pair of vectors chosen would be those that give the minimum magnitude of the difference between those heading vectors. However, for *N* data points, this leads to $\frac{1}{2}(N-1)(N-2)$ comparisons which can be excessive. In the Global 1 method, to speed up the clustering process, we use a method based on ordering the individual components of the heading vectors. This method is described in detail in [25], and this is repeated below for convenience.

Steps 1 to 3 for Global 1 are the same as for the MHC method leading up to Table 2. Subsequent steps of the clustering algorithm are as follows:

Step 4: Take absolute values of each component of the normalised heading vector in Table 2: $|\hat{r}_i[n]|$

For the example in Table 2, the magnitudes of the heading components are as shown in Table 4.

| Normalised Heading Number, $n$ | $|\hat{r}_1[n]|$ | $|\hat{r}_2[n]|$ |
|---|---|---|
| 1 | 0.4472 | 0.8944 |
| 2 | 0.5547 | 0.8321 |
| 3 | 0.4472 | 0.8944 |
| 4 | 0.4472 | 0.8944 |
| 5 | 0.8575 | 0.5145 |
| 6 | 0.4472 | 0.8944 |
| 7 | 0.5547 | 0.8321 |
| 8 | 0.7071 | 0.7071 |
| 9 | 0.5547 | 0.8321 |
| 10 | 0.4472 | 0.8944 |

*Table 4 – Magnitude of Normalised Heading Components*

Step 5: For each component $i$, sort absolute values of normalised heading components in ascending order

$$\{\hat{r}^{sort}_i[m]\} = \text{sort}\{|\hat{r}_i[n]|\} \qquad (54)$$

Note the following mapping between $n$ and $m$ for each component $i$

$$f_{1i}[m] = n \qquad (55)$$

The sorted headings and the array $\{f_{1i}[m]\}$ for the above diagram are shown in Table 5:

| Sorted Normalised Heading Number, $m$ | $\lvert \hat{r}^{sort}_1[m] \rvert$ | $f_{11}[m]$ | $\lvert \hat{r}^{sort}_2[m] \rvert$ | $f_{12}[m]$ |
|---|---|---|---|---|
| 1 | 0.4472 | 1 | 0.5145 | 5 |
| 2 | 0.4472 | 3 | 0.7071 | 8 |
| 3 | 0.4472 | 4 | 0.8321 | 2 |
| 4 | 0.4472 | 6 | 0.8321 | 7 |
| 5 | 0.4472 | 10 | 0.8321 | 9 |
| 6 | 0.5547 | 2 | 0.8944 | 1 |
| 7 | 0.5547 | 7 | 0.8944 | 3 |
| 8 | 0.5547 | 9 | 0.8994 | 4 |
| 9 | 0.7071 | 8 | 0.8944 | 6 |
| 10 | 0.8575 | 5 | 0.8944 | 10 |

*Table 5 – Sorted Heading Components*

Step 6: Cluster heading components by looking at the differences between adjacent values of $\hat{r}^{sort}_i[m]$ given by $\lvert \hat{r}^{sort}_i[m] - \hat{r}^{sort}_i[m-1] \rvert$.

Choose a threshold $\varepsilon$. Then populate a matrix **C** as follows:

if $\lvert \hat{r}^{sort}_i[m] - \hat{r}^{sort}_i[m-1] \rvert < \varepsilon$

$C_i[m] = 1$

    else

$C_i[m] = 0$

Let **C** be a matrix for which $C_i[m]$ is the element in the $m^{th}$ row and $i^{th}$ column.

In the above example there is assumed to be no noise. In practice, there will be noise and that is the reason why we allow the difference between adjacent values of $\hat{r}^{sort}_i[m]$ to be less than some non-zero threshold $\varepsilon$; we will discuss later how to choose $\varepsilon$.

For our example, the values of $\{C_i[m]\}$ and $f_{1i}$ are shown in Table 6.

| Sorted Normalised Heading Number, $m$ | $C_1[m]$ | $f_{11}[m]$ | $C_2[m]$ | $f_{12}[m]$ |
|---|---|---|---|---|
| 1 | 0 | 1 | 0 | 5 |
| 2 | 1 | 3 | 0 | 8 |
| 3 | 1 | 4 | 0 | 2 |
| 4 | 1 | 6 | 1 | 7 |
| 5 | 1 | 10 | 1 | 9 |
| 6 | 0 | 2 | 0 | 1 |
| 7 | 1 | 7 | 1 | 3 |
| 8 | 1 | 9 | 1 | 4 |
| 9 | 0 | 8 | 1 | 6 |
| 10 | 0 | 5 | 1 | 10 |

*Table 6 – C values*

Step 7: Look for column $j$ in $\{C_i[m]\}$ with largest number of adjacent values of 1's:

*Notes*

(i) In Table 6 it can be seen that there are two such clusters of 1's: $C_1[m]$ for $2 \leq m \leq 5$ and $C_2[m]$ for $7 \leq m \leq 10$. In this case, the software picks up the cluster of $C_1$ values, but the same final result will be obtained if the other cluster is picked first.

(ii) Note that $C_1[1] = C_1[6] = C_2[3] = C_2[6] = 0$. The reasons for putting zeros at these points in the Table is to separate clusters of 1's corresponding to different headings; for example, if $C_1[6]$ was put equal to 1 then there would be a continuous cluster of 1's from $C_1[2]$ to $C_1[8]$ implying that all these components come from the same heading which, clearly, they do not.

Step 8: Now that we have identified a clustering of a heading component in Table 6, we now need to associate these components to the time ordered components shown in Table 4.  This is where the values of $f_{11}[m]$ and $f_{12}[m]$ in Table 6 are used.  In this Table, the following normalised sorted heading values are clustered together $|\hat{r}^{sort}_1[2]|, |\hat{r}^{sort}_1[3]|, |\hat{r}^{sort}_1[4]|$ and $|\hat{r}^{sort}_1[5]|$. Following on from Note (ii) in Step (7) above we should also include $|\hat{r}^{sort}_1[1]|$ in the clustering. Using the values of $f_{11}[m]$ in this Table, these sorted heading components correspond to the following unsorted heading components: $|\hat{r}_1[1]|, |\hat{r}_1[3]|, |\hat{r}_1[4]|, |\hat{r}_1[6]|$ and $|\hat{r}_1[10]|$. We now define a matrix $\mathbf{C^U}$, where the element in the $n^{th}$ row and $i^{th}$ column is $C^U_i[n]$; we fill in 1's in column 1 at rows 1,3,4,6 and 10 indicating that the clustering of the first heading component has taken place, as shown in Table 7.

| Normalised Heading Number, $n$ | $C^U_1[n]$ | $C^U_2[n]$ |
|---|---|---|
| 1 | 1 | |
| 2 | 0 | |
| 3 | 1 | |
| 4 | 1 | |
| 5 | 0 | |
| 6 | 1 | |
| 7 | 0 | |
| 8 | 0 | |
| 9 | 0 | |
| 10 | 1 | |

*Table 7 – Time ordered values for $C^U_1[n]$*

Step 9:   Now, we have found that the largest cluster for component 1 corresponds to the original sample numbers 1, 3, 4, 6 and 10.  We now need to determine how many heading values at these sample numbers for component 2 are also in a cluster.  To determine this, look at Table 6.  We can see that the values of $C_2[n]$ corresponding to these time points are 0,1,1,1,1.  However, for the reason stated in Step 7(ii), we need to put $C_2[6] = 1$ as this is part of the same cluster corresponding to $C_2^U[1] = 1$. Hence the second column of the above table can be filled in as follows:

| Normalised Heading Number, $n$ | $C^U{}_1[n]$ | $C^U{}_2[n]$ |
|---|---|---|
| 1 | 1 | 1 |
| 2 | 0 | 0 |
| 3 | 1 | 1 |
| 4 | 1 | 1 |
| 5 | 0 | 0 |
| 6 | 1 | 1 |
| 7 | 0 | 0 |
| 8 | 0 | 0 |
| 9 | 0 | 0 |
| 10 | 1 | 1 |

*Table 8 – Time ordered values for $C^U{}_1[n]$ and $C^U{}_2[n]$*

Step 10: In Table 8, we are looking for rows where all elements $\{C^U{}_i\}$ are 1; in this case both components are part of a cluster. This can be achieved by performing a logical AND of the elements of each row according to

$$D[p] = \text{AND}\ (C^U{}_1[p], C^U{}_2[p]) \tag{56}$$

The following is a Table showing $\{D[n]\}$ values for the example:

| Normalised Heading Number, $n$ | $D[p]$ |
|---|---|
| 1 | 1 |
| 2 | 0 |
| 3 | 1 |
| 4 | 1 |
| 5 | 0 |
| 6 | 1 |
| 7 | 0 |
| 8 | 0 |
| 9 | 0 |
| 10 | 1 |

*Table 9 – D[p]*

Step 11:      Each row, *p*, where *D[p]* = 1 corresponds to a heading in the same cluster; for the above example, the following headings form a cluster:

$\hat{r}[1], \hat{r}[3], \hat{r}[4], \hat{r}[6]$ and $\hat{r}[10]$

which agrees with Table 4.

*Estimation of Heading Vector*

Let $\boldsymbol{v}^a = (v_1{}^a, v_2{}^a, \ldots, v_N{}^a)$ denote the components of the actual *N*-dimensional non-normalised velocity vector. Suppose that the clustering algorithm has been carried out on each component and let us look at velocity component *i*, $v_i{}^a$. Let us assume that there are *J* velocity vectors within a cluster.

Let $\hat{\boldsymbol{r}}^e = \{\hat{r}_i{}^e\}$ denote the estimate of the heading from a particular cluster of headings. We now need to estimate $\{\hat{r}_i{}^e\}$ from the set of velocity vectors that have been found from the clustering method: $\{v_i'[1]\}, \{v_i'[2]\}, \ldots, \{v_i'[J]\}$, where $v_i'[n]$ is the *i*[th] component of the *n*[th] velocity vector in a cluster.

Each velocity component will be affected by noise. One possibility to determine the $i^{th}$ component of the estimated heading, $\hat{r}_i^e$, is to perform a direct average over $j$ of $v_i'[j]$. However this is not optimal because noise will affect smaller magnitude velocities more than those with larger magnitude. Hence any estimator should take this into account by putting more weight on larger magnitude velocity components than the smaller magnitude velocity components when averaging over all components in a cluster, because the signal-to-noise ratio for the former components is larger. This problem is addressed in [34] for the averaging of evoked potentials, where it is shown that the estimate, $\tilde{V}_i$, of the $i^{th}$ velocity component is given by

$$\tilde{V}_i = \frac{\sum_{j=1}^{J} M[j] \cdot v_i'[j]}{\sum_{j=1}^{J} (M[j])^2} \tag{57}$$

where

$$M[j] = \sqrt{\sum_{i=1}^{N} \{v_i'[j]\}^2} \tag{58}$$

When $M[j] = 1$ for all $j$, this reduces to a straight average over all headings.

The above processing is applied to each velocity component $i$ (=1,…,$N$) so that the estimate of the velocity vector becomes:

$$\widetilde{V} = (\tilde{V}_1, \tilde{V}_2, \ldots, \tilde{V}_N) \tag{59}$$

The estimate of the normalised heading vector is then

$$\widehat{R}_e = \frac{\widetilde{V}}{|\widetilde{V}|} \tag{60}$$

$\widehat{R}_e$ is then used in Equation (25) in place of $\widehat{R}_p$.

*Input Parameters*

We now need to consider the choice of ε in Step 6 above. This parameter is used to determine if two heading components are associated with the same source. In the regions where one source is on its own, the ideal value for ε is zero. In practice, because of the effects of noise and other sources, a non-zero value for ε should be chosen. In Reference [25], this point is discussed further where it is argued that a reasonable choice for ε is

$$\varepsilon = \alpha \frac{1}{M} \tag{61}$$

where $\alpha \leq 1$.

If α is chosen to be too small, then associations between heading components belonging to the same source will not occur; if too large, then too many false associations will occur between headings that are not from the same source. According to Step 10 above, a cluster is only declared if an association is found between all components of the heading; hence, randomly associated heading components will tend to AND to zero.

Extensive simulations have been carried out in [25] to optimise the parameter α in (61) and it has been found that a good compromise value to use is α = 1; this is used in the simulations and data analysis carried out in this paper.

*Simulations*

The MHC is applied to the mixture of sources illustrated in Figure 2. In Figure 6, the detected clusters of headings are shown at each iteration, which correspond to the occurrence of the peaks in Figure 4 and the estimate of the sources are shown in Figure 7.
Note that at the first iteration, two clusters of headings are detected, one between sample numbers 0 and 50 corresponding to the co-incidence of peaks between the two sources and a larger cluster of headings around sample point 275 corresponding to Source 1; as this cluster has the larger number of headings, these are used to detect Source 1. Note that the headings within the cluster in this case are identical because of the absence of noise in the data. At the second iteration, the clusters of headings correspond to the peaks of Source 2
Note that, as for MHC, there is a contribution from Source 2 to the estimate of Source 1.

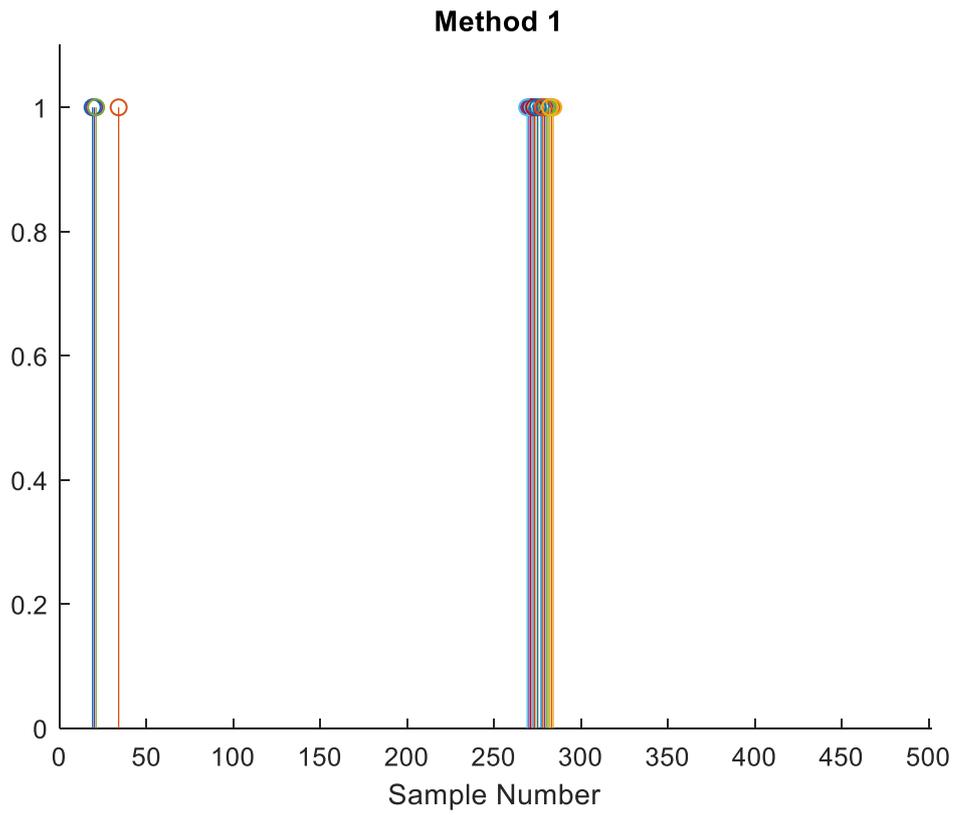

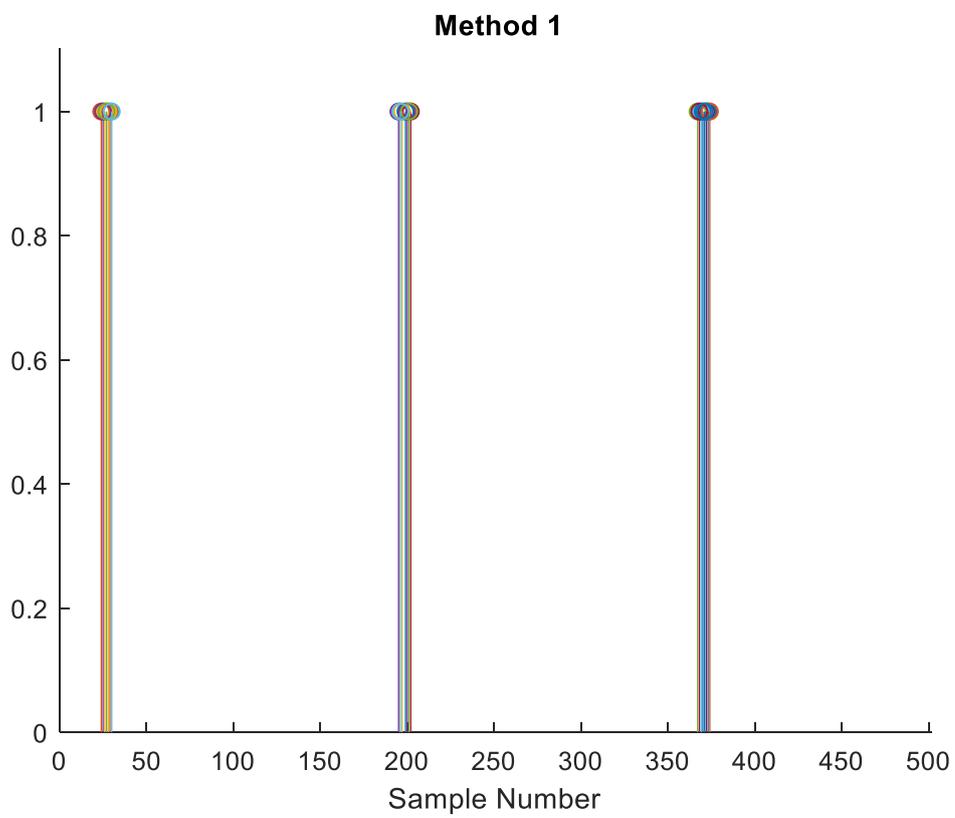

Figure 6 – detected clusters: Top figure: first iteration, detecting Source 1, Bottom figure: second iteration detecting Source 2

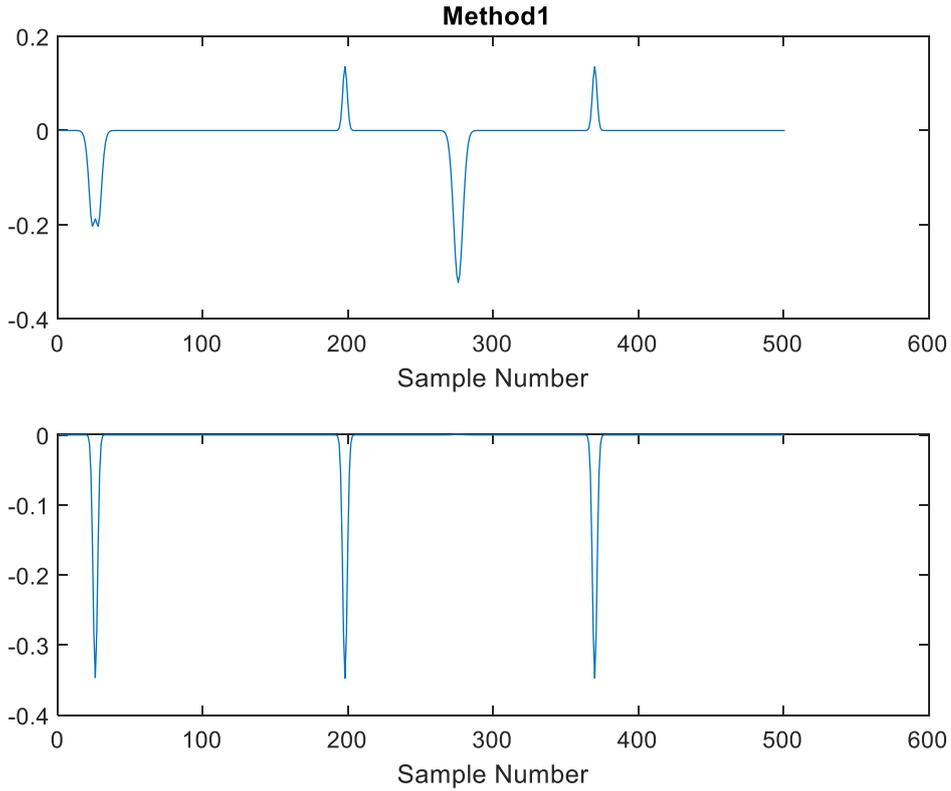

Figure 7 – Estimates of Source 1 (top) and Source 2 (bottom)

The potential advantage of Global 1 over the MHC method is that averaging over heading components reduces the effects of noise, although outliers in a group may contribute biases. The association between similar headings can take place at different parts of the input data not necessarily at adjacent time points.

The disadvantage of Global 1 compared to MHC is the requirement to define an additional threshold, (61). In addition, this method requires sorting and resorting of heading components, which is not required by the MHC.

### 3.2.2    Global 2

As mentioned at the beginning of Section 3.2.1, directly comparing pairs of headings is a computationally intensive task. Global 1 is a such a method to reduce the computational load, although it requires the use of a threshold parameter ε in (61); however, there is an alternative to Global 1, where this parameter is not needed – this method we call Global 2. Like Global 1, this method is based on sorting of heading components. Global 2 initially follows Steps 1 to 5 for Global 1. To illustrate the principle of the proposed method, let us look at the example to explain Global 2 starting at Step 5 for Global 1. At this step, the magnitude of each heading component has been sorted in ascending order, for example Table 5 as shown below:

| Sorted Normalised Heading Number, $m$ | $|\hat{r}^{sort}_1[m]|$ | $f_{11}[m]$ | $|\hat{r}^{sort}_2[m]|$ | $f_{12}[m]$ |
|---|---|---|---|---|
| 1 | 0.4472 | 1 | 0.5145 | 5 |
| 2 | 0.4472 | 3 | 0.7071 | 8 |
| 3 | 0.4472 | 4 | 0.8321 | 2 |
| 4 | 0.4472 | 6 | 0.8321 | 7 |
| 5 | 0.4472 | 10 | 0.8321 | 9 |
| 6 | 0.5547 | 2 | 0.8944 | 1 |
| 7 | 0.5547 | 7 | 0.8944 | 3 |
| 8 | 0.5547 | 9 | 0.8994 | 4 |
| 9 | 0.7071 | 8 | 0.8944 | 6 |
| 10 | 0.8575 | 5 | 0.8944 | 10 |

*Table 5 – Sorted Heading Components*

Step 6a: Compute Differences between Adjacent Sorted Heading Values

In Global 2, the differences between adjacent sorted heading components are computed and these are shown in Table 10 below:

| Sorted Normalised Heading Number, $m$ | $\Delta|\hat{r}^{sort}_1[m]|$ | $f_{11}[m]$ | $\Delta|\hat{r}^{sort}_2[m]|$ | $f_{12}[m]$ |
|---|---|---|---|---|
| 1 | x | 1 | x | 5 |
| 2 | 0 | 3 | 0.1926 | 8 |
| 3 | 0 | 4 | 0.1250 | 2 |
| 4 | 0 | 6 | 0 | 7 |
| 5 | 0 | 10 | 0 | 9 |
| 6 | 0.1075 | 2 | 0.0623 | 1 |
| 7 | 0 | 7 | 0 | 3 |
| 8 | 0 | 9 | 0 | 4 |
| 9 | 0.1524 | 8 | 0 | 6 |
| 10 | 0.1504 | 5 | 0 | 10 |

*Table 10 - differences between adjacent sorted heading components*

Note that the difference between heading components is undefined for *m* = 1 and this is indicated by an "x" in the above Table.

Let us look at an example. Sorted heading component 1 at *m*=6 in Table5 $|\hat{r}_1^{sort}[6]|$=0.5547 is different to the corresponding component at *m*=5 $|\hat{r}_1^{sort}[5]| = 0.4472$ and this is reflected in a non-zero value of the difference between the two $\Delta|\hat{r}_1^{sort}[6]| = 0.1075$ at *m*=6 in Table 10.

However, in Table 5 it can be seen that sorted heading component 1 at *m*=5, $|\hat{r}_1^{sort}[5]| = 0.4472$ is the same as that at *m*=4 $|\hat{r}_1^{sort}[4]| = 0.4472$ and this is reflected by a zero value in the corresponding entry in Table 10: $\Delta|\hat{r}_1^{sort}[5]| = 0$.

Step 7a:     Reorder the heading differences in order of time.

In this step, we reorder the heading component differences in order of time again. Let $\Delta_1[n]$ and $\Delta_2[n]$ be defined as:

$$\Delta_1[f_{11}[m]] = \Delta|\hat{r}_1^{sort}[m]| \tag{62}$$

and

$$\Delta_2[f_{12}[m]] = \Delta|\hat{r}_2^{sort}[m]| \tag{63}$$

$\Delta_1[n]$ and $\Delta_2[n]$ represent a reordering in time of the heading differences.

| Unsorted Normalised Heading Number, *n* | $\Delta_1[n]$ | $\Delta_2[n]$ |
|---|---|---|
| 1 | x | 0.0623 |
| 2 | 0.1075 | 0.1250 |
| 3 | 0 | 0 |
| 4 | 0 | 0 |
| 5 | 0.1504 | x |
| 6 | 0 | 0 |
| 7 | 0 | 0 |
| 8 | 0.1524 | 0.1926 |
| 9 | 0 | 0 |
| 10 | 0 | 0 |

*Table 11 - differences between adjacent sorted heading components reordered in time*

For example, in Table 10, when $m = 2$, $f_{11}[2] = 3$, so from (62), $\Delta_1[3] = \Delta|\hat{r}_1^{sort}[2]| = 0$. Also, at $m = 2$, $f_{12}[2] = 8$ so that from (63) $\Delta_2[8] = \Delta|\hat{r}_2^{sort}[2]| = 0.1926$.

We can now interpret Table 11 as follows.

At time point $n = 8$, the differences $\Delta_1[n]$ and $\Delta_2[n]$ are relatively large in Table 11, which means that the heading components at this time point do not associate with the previous components in the sorted Table 10.

At time point $n=3$, the differences $\Delta_1[n] = \Delta_2[n] = 0$ in Table 11, which means that this heading does associate with another heading; note that the two headings that are being associated *need not be adjacent in time*; they could occur at any points in the data interval where one source exists on its own.

Step 8a:   Determine magnitude of the differences in Table 11:

$$E[n] = \sqrt{(\Delta_1[n])^2 + (\Delta_2[n])^2} \qquad (64)$$

| Unsorted Normalised Heading Number, $m$ | $E[n]$ |
|---|---|
| 1 | x |
| 2 | 0.1649 |
| 3 | 0 |
| 4 | 0 |
| 5 | x |
| 6 | 0 |
| 7 | 0 |
| 8 | 0.2456 |
| 9 | 0 |
| 10 | 0 |

*Table 12 – E[n], Equation (64)*

Step 9a:   Choose heading number $n$ with the minimum $E[n]$. In the above example, Table 11, this can be heading numbers 3,4,6,7,9 or 10. The heading at $n = 2$ is not detected because this method is based on differences in headings rather than the headings themselves, so we lose a heading – this is not a problem in practice, particularly for longer data records. Note that each of the above chosen headings is a subset of either (1,3,4,6,10) or (2,7,9) so a member of either group would have been chosen. The above example is idealised; in the more practical situation where noise is present or where sparsity is not perfect then there will in general be one heading that has the smallest $E[n]$, which will hopefully be the least affected by noise.

The advantage of Global 2 over Global 1 is that there is no need to define a threshold, $\varepsilon$, to group headings; we are just looking for minimum difference in magnitudes of adjacent headings when the components are sorted in ascending order.

The disadvantages of Global 2 are

- Like Global 1, the sorting and resorting of headings can take time.
- No clustering of headings takes place and so there is no smoothing of noise by averaging headings within a cluster as is done for Global 1 (57). To compensate for this, to minimise the effects of noise, we can choose the velocity vectors with the largest magnitudes by increasing threshold $v^{th}$ in (61); the headings will then correspond to the velocity vectors with the largest magnitude, which should be least affected by noise .

Let us look again at the example of mixtures illustrated in Figure 2. The detected headings for the sources when using Global 2 are shown in Figure 8. The detected headings correspond to the peak positions in the source signals, with Source 2 being detected before Source 1. Note that, like MHC, only one heading per component is detected. However, as for MHC and Global 1 there is a contamination of the estimate of Source 2 by Source 1 (Figure 9, top) due to correlation between the source signals.

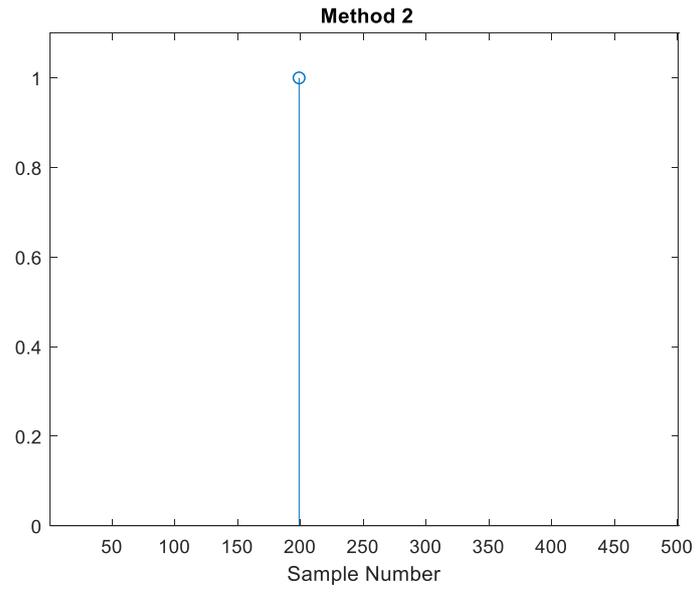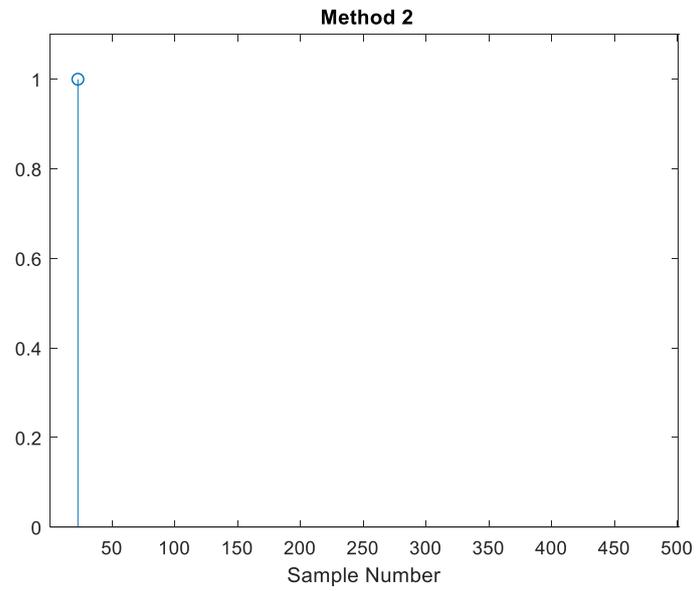

Figure 8 - Samples where Sources are Detected using Global 2

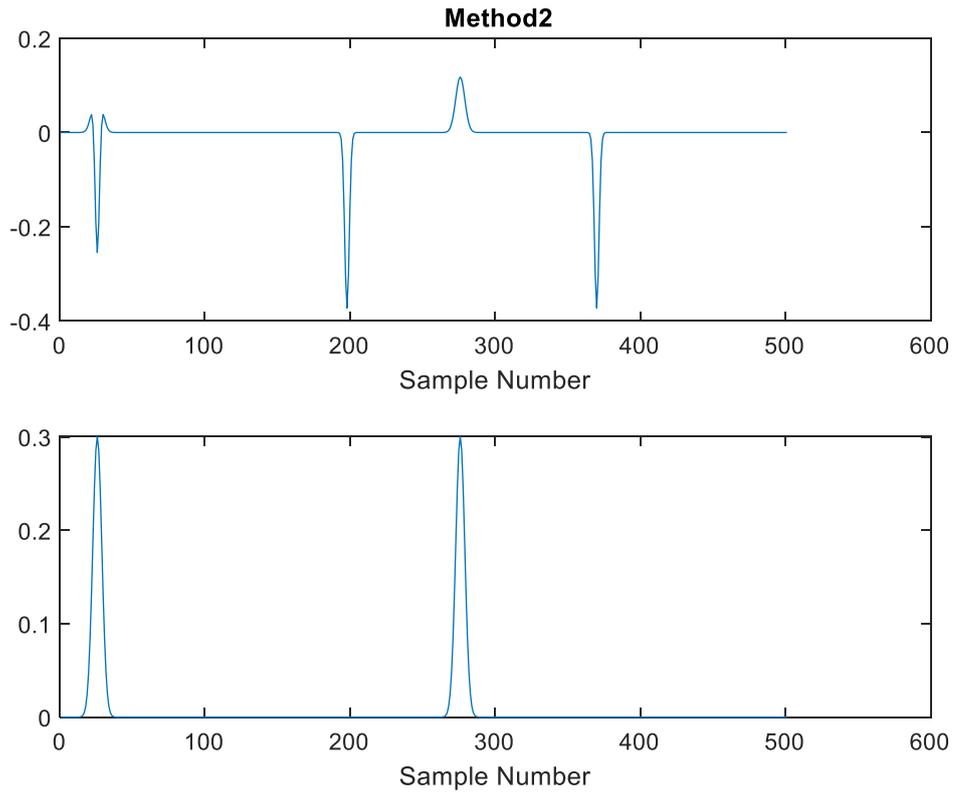

Figure 9 – Estimated Sources. Top: first iteration, Bottom: second iteration.

### 3.2.3 Fast ICA

In this paper, we will be comparing the MHC, Global 1 and Global 2 methods with FastICA. If we apply FastICA, using the deflation method with Gaussian non-linearity, to the mixed data in Figure 2, we obtain the estimates of the sources. shown in Figure 10.

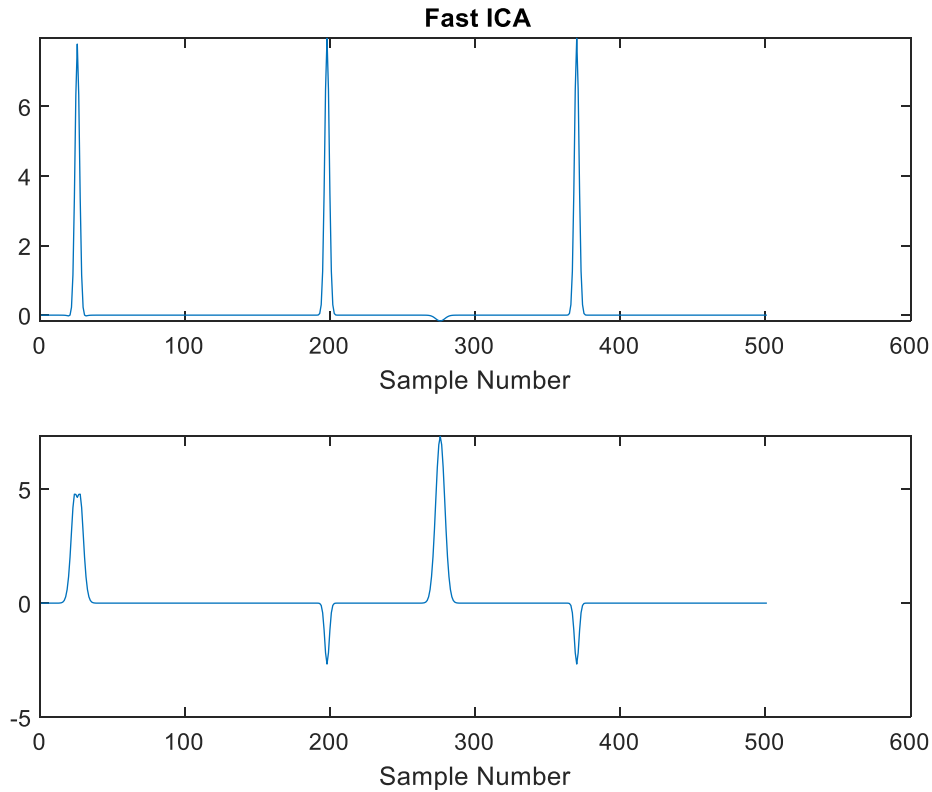

Figure 10 – Estimate of Two Sources using FastICA

In this case, the estimates of Sources 1 and 2 are both contaminated by the other source. The reason here is that FastICA processes the whole data, whilst MHC, Global Method 1 and Global Method 2 use information on the sparse sections and do not process the whole data; as shown in Section 2.2, this means that in theory the final source estimated is uncontaminated by any other sources.

## 4. Results

The method Global 2 is now tested on mixtures of three types of sources: non-overlapping sources, partially overlapping sources and completely overlapping sources. The method is also applied to practical ECG signals taken from an expectant mother. A comparison of the results is made between the following methods: FastICA, MHC, Global 1 and Global 2. The deflation approach is used when applying the FastICA method; as we will see below, the method can be sensitive to the non-linearity used depending on the sources.

The figure of merit used, for the simulations, to compare the estimated and actual sources is the RMS error as a function of sample point $n$.

- Firstly, we need to normalise the estimates and actual sources to take into account scaling differences between the estimates and the sources.

  If the unscaled samples of a general, signal are $\{x[1], x[2], \ldots, x[M]\}$, then, we normalise the signal so that the normalised signal samples are given by $\{x'[1], x'[2], \ldots, x'[M]\}$, where

  $$x'[n] = \frac{x[n]}{\sqrt{\sum_{i=1}^{M}(x[i])^2}} \qquad (65)$$

  Note that this normalisation was used in reference [25]; however it was incorrectly described as normalising the signal to have an rms value of 1, which would have involved an extra $1/\sqrt{M}$ factor in the denominator of Equation (65).

  Let the normalised actual source and the normalised estimate of the source at time point $n$ for Monte Carlo run $q$ be $s_r[n]$ and $\tilde{s}_s[n]$ respectively.

- Next, we correlate each source with each estimate and put the correlation coefficients in a matrix, with the rows representing different sources and the columns representing different estimates; let the correlation between source $r$ and estimate $s$ be $c_{rs}{}^q$ where $q$ denotes the Monte Carlo run number.

- In each row of the correlation matrix, we choose the entry with maximum magnitude of the correlation value to associate a particular source with a particular estimate.

- We use the sign of the correlation value to detect inversion of the estimates and compute the difference from

  $$\varepsilon_s{}^q[n] = s_r[n] - \tilde{s}_s[n] \qquad \text{if } c_{rs}{}^q > 0$$
  $$= s_r[n] + \tilde{s}_s[n] \qquad \text{if } c_{rs}{}^q < 0 \qquad (66)$$

- Determine rms estimation error as a function of time from Equation (30) of reference [25]:

  $$RMS[n] = \sqrt{\frac{1}{Q}\sum_{q=1}^{Q}\left(\varepsilon_s^q[n]\right)^2} \qquad (67)$$

  where $Q$ is the number of Monte Carlo runs.

  In [25] the maximum of RMS[$n$] was calculated along with the rms error averaged over all sample points. In this this paper, we display rms errors as a function of sample number as this gives more information on the effects of contamination of other sources on the estimates of a particular source; this figure of merit was not displayed in [25] so will give additional information about the comparative performances of the four methods under test.

## 4.1 Completely Sparse Sources

The first example we will look at is the case of a mixture of two completely sparse sources – in this case, each source is non-zero for non-overlapping segments of time. The pure sources are shown in Figure 11 (note the different y-scales) and the mixtures are calculated using the randomly chosen mixing matrix:

$$\mathbf{A} = \begin{pmatrix} 1.3 & 2 \\ 1 & 3 \end{pmatrix}$$

Firstly, we will look at the case where the data mixtures are noise-free. According to the analysis in Section (2.2), the MHC, Global 1 and Global 2 methods should all extract each source separately to within a scaling constant. To save space, this is shown for the Global 2 method only in Figure 12, with the threshold parameter $v^{th} = 0.1$ in Equation (46).

The corresponding results for the Fast ICA method, (deflation method, Gaussian non-linearity) is shown in Figure 13, where it is seen that the estimate of each source estimate is contaminated by the other source; this contamination comes from the subtraction of the mean when whitening the data which introduces correlation between the sources prior to applying the FastICA,

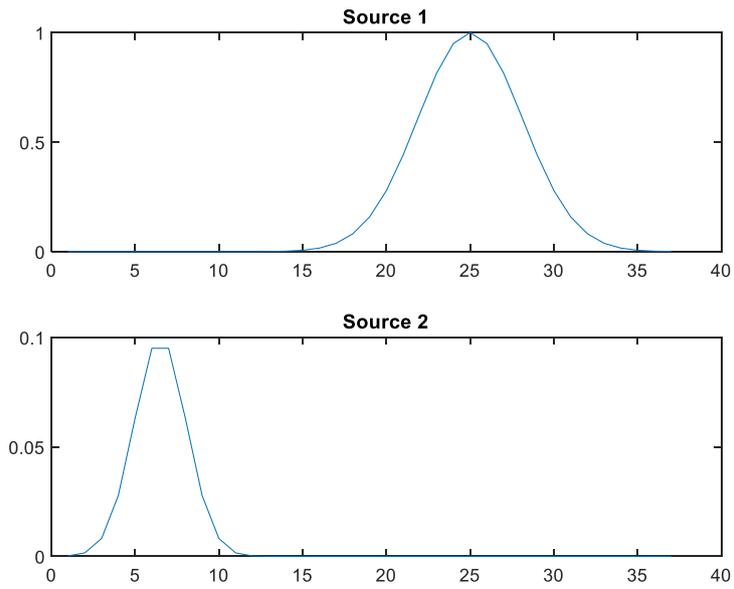

Figure 11 – Purely Sparse Sources

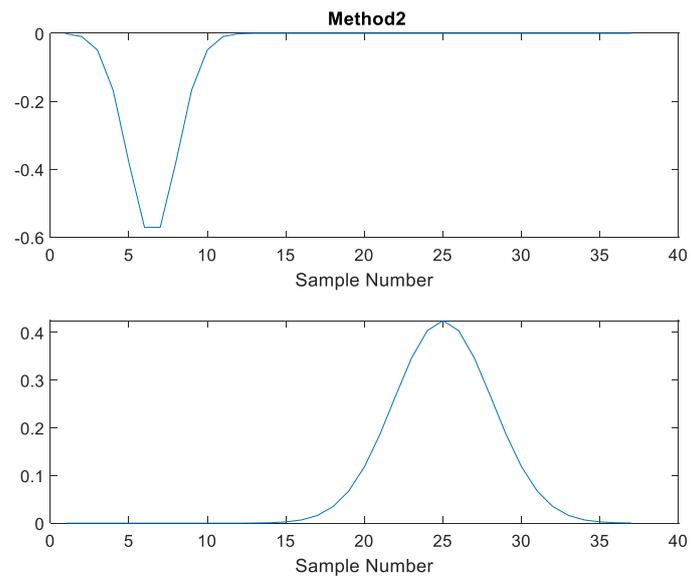

Figure 12 – Sources Estimated using Global 2

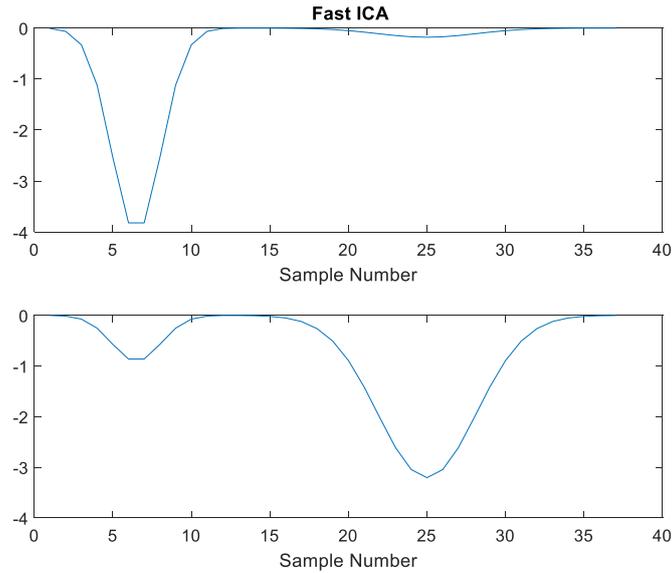

Figure 13 – Sources Estimated using FastICA

We now compare the robustness to noise of Global 2 with the Global 1, MHC and FastICA methods. The noise sd = 0.005, which is 2.5 % of the maximum value of the smaller source in the mixed data. the RMS errors as a function of sample number is calculated according to (67), and Q=10 000 Monte Carlo runs are carried out.

In subsequent figures, for each source, two plots are displayed (a) the RMS errors as a function of sample number for each of the four methods that have been applied (b) as for (a), the RMS errors are displayed and, in addition, the clean source signals, (shown in magenta), which have been normalised according to Equation (65) are plotted ; comparing RMS[$n$] with this normalised source will show the significance of the errors across the duration of the source and also indicate any contaminations of one source estimate from the other. The threshold values in (46) for the Global 1, Global 2 and MHC methods are respectively, 0.1, 0.9 and 0.8.

**Source 1(a)**

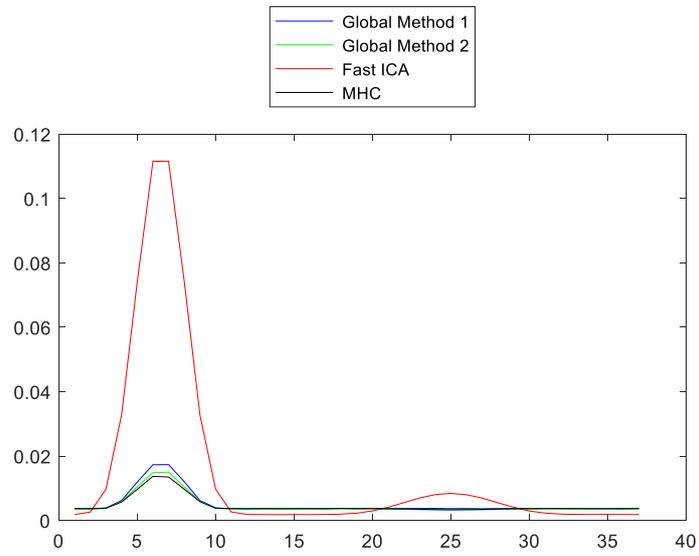

**Source 1(b)**

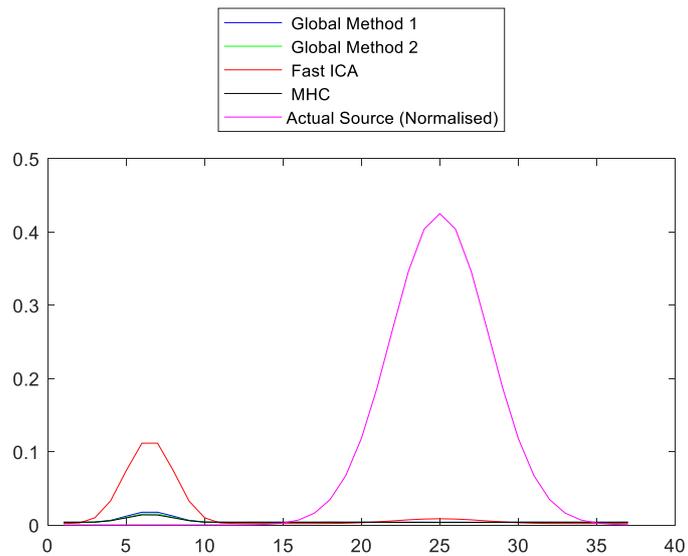

Figure 14 (Source 1) – Comparison of RMS Errors as a function of sample number for completely sparse sources (a) rms errors for Source 1 (b) rms errors for Source 1 along with clean signal.

**Source 2(a)**

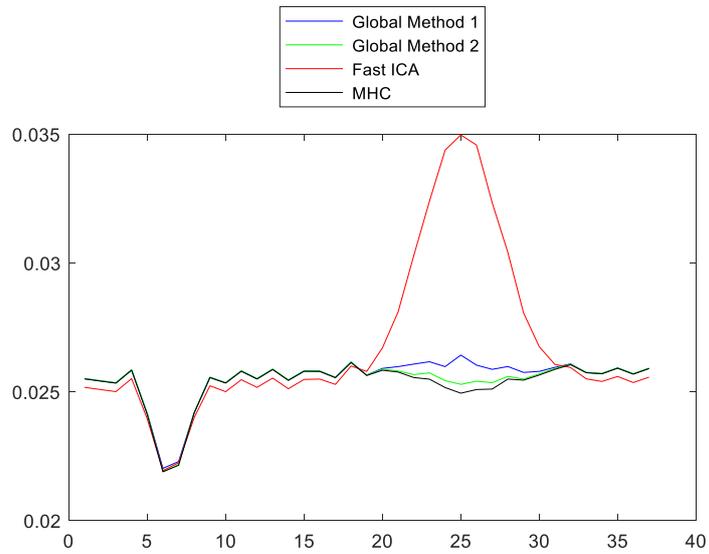

**Source 2(b)**

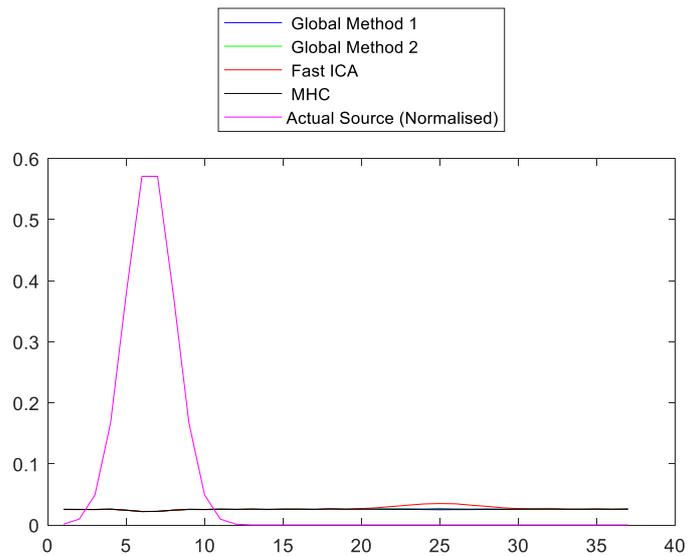

Figure 14 (Source 2) – Comparison of RMS Errors as a function of sample number for completely sparse sources (a) rms errors for Source 2 (b) rms errors for Source 2 along with clean signal.

Looking at Figure 14 (Source 1) and Figure 14 (Source 2), there are similar performances between the MHC, Global 1 and Global 2 methods. For the estimate of Source 1, all four methods have significant rms errors between sample numbers 3 and 10 because of contamination from Source 2. The reason for this is that the presence of noise means that the principal heading vector in Equation (24) will be in error which will pick up contributions from the other source when used in Equation (25). This is the case for all four methods. However, an additional problem for the FastICA is the introduction of correlation between the sources after subtracting the mean when data whitening; for that reason, for FastICA the contribution from Source 2 to the estimate of Source 1 is the most significant for the four methods under test. Looking at the estimates for Source 2, FastICA performs the worst of the four methods. It has been found that, as expected, Global 2 has best robustness to noise for increasing threshold value $v^{th}$ in (46). Note that for Global 1, if $v^{th}$ is chosen to be too large then not enough headings can be found to form a cluster, and the method breaks down.

## 4.2 Partially Sparse Sources

We have already looked at the analysis of data consisting of a mixture of two partially overlapping sources in Sections 1 and 3.

In this section, we compare the performances of the algorithms when there is noise added to the data. We will look at the case when the noise has a standard deviation of 0.05 which is 5 % of the maximum value of the smaller source in the mixed data. The RMS errors are calculated for 10000 Monte Carlo runs.

The presence of noise in the data will affect the estimation of principal vectors in phase space and hence there will be contamination of one source estimate by the other.

For Global 1, the best results are obtained when the threshold parameter $v^{th} = 0.4$. It should be noted that, as for the case of purely sparse sources (Section 4.1), if $v^{th}$ is chosen to be too large then no clusters will form, and the algorithm will break down. For the MHC, best results are obtained with $v^{th} = 0.7$ and the same optimal value of $v^{th}$ is found for Global 2.

The results for RMS[$n$] are shown in Figure 15 for Sources 1 and 2.

**Source 1(a)**

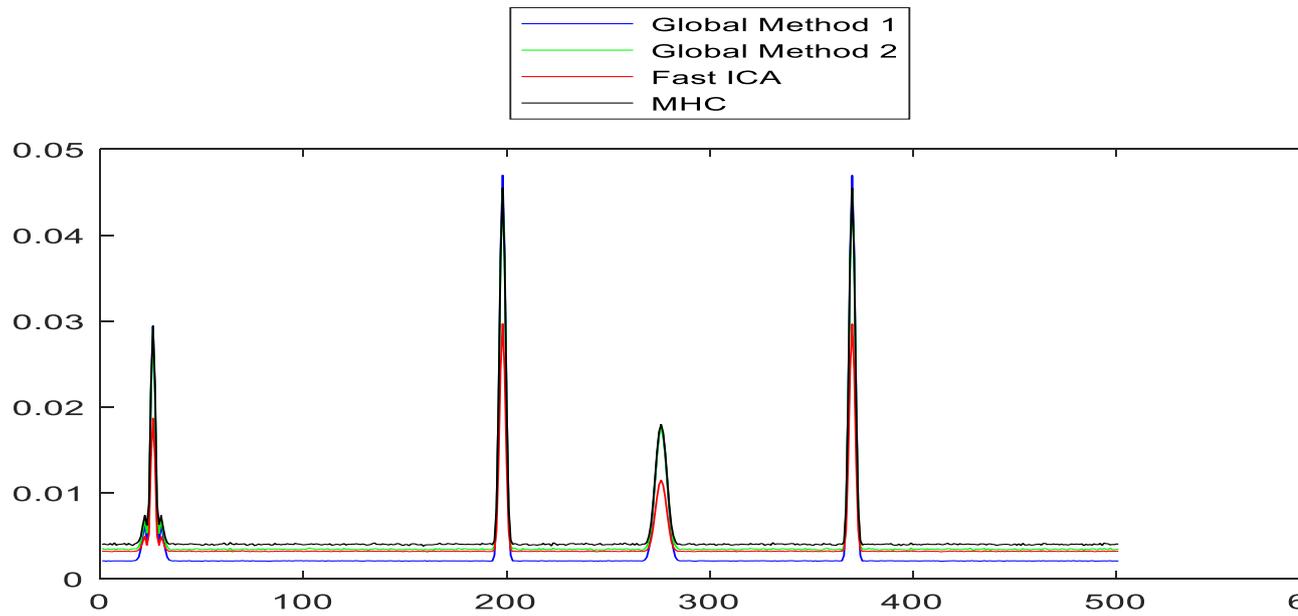

**Source 1(b)**

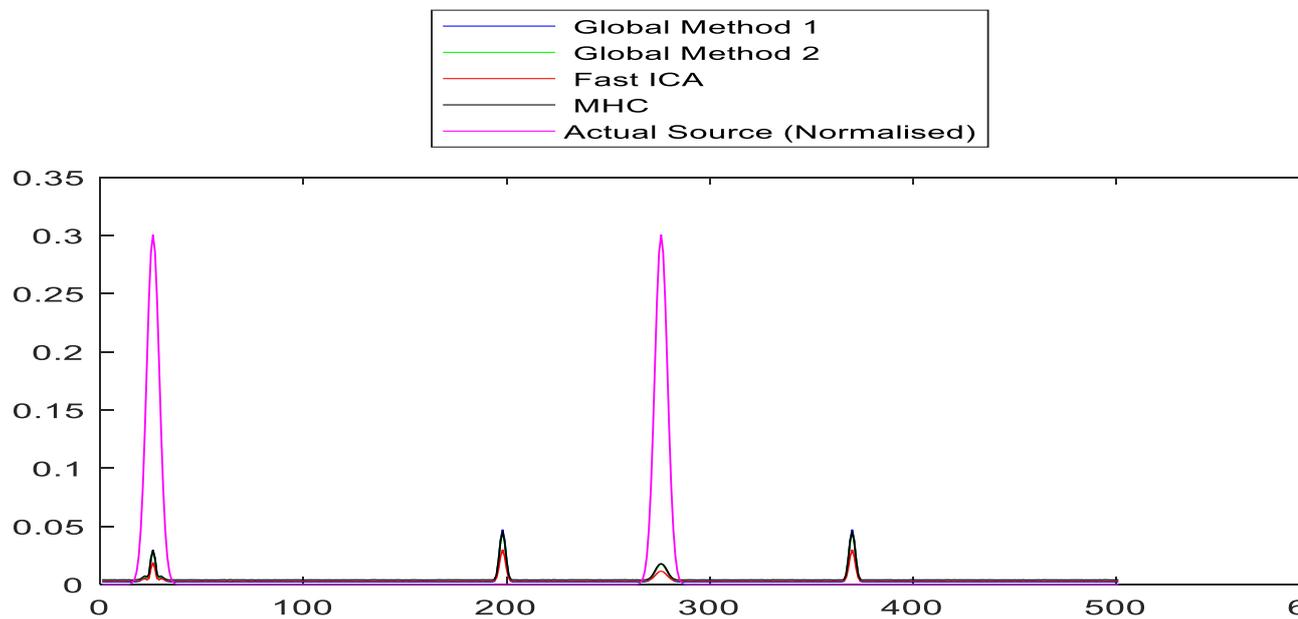

Figure 15 (Source 1) – Comparison of RMS Errors as a function of sample number for Source 1 (a) rms errors for Source 1 (b) rms errors for Source 1 along with clean signal.

**Source 2(a)**

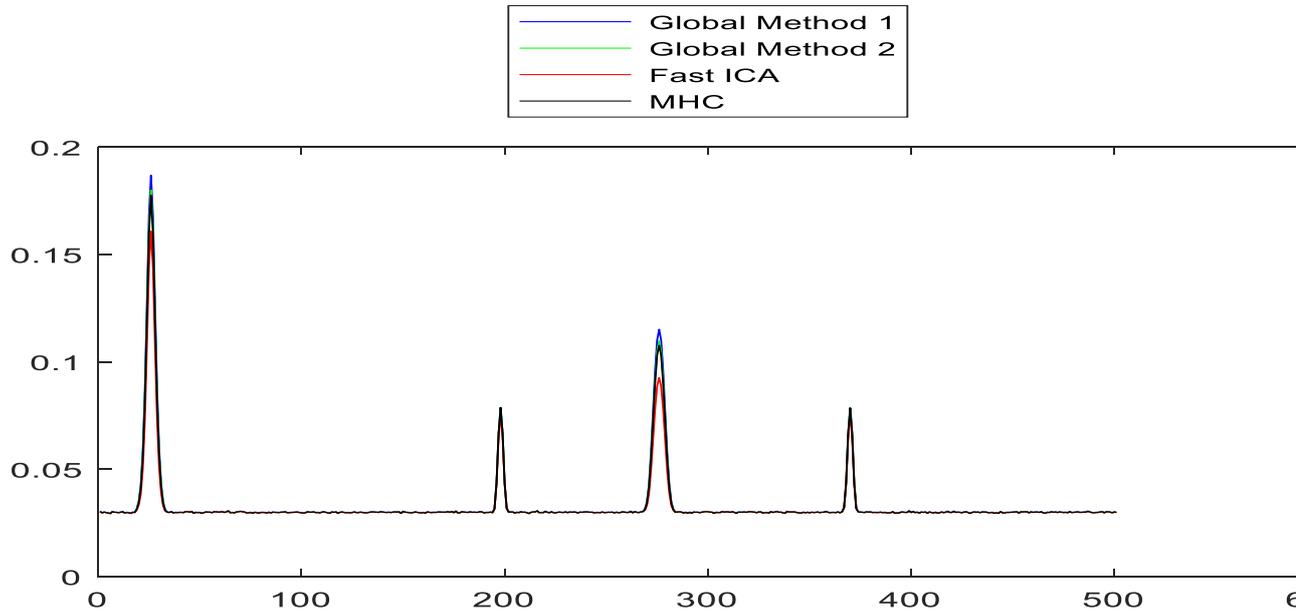

**Source 2(b)**

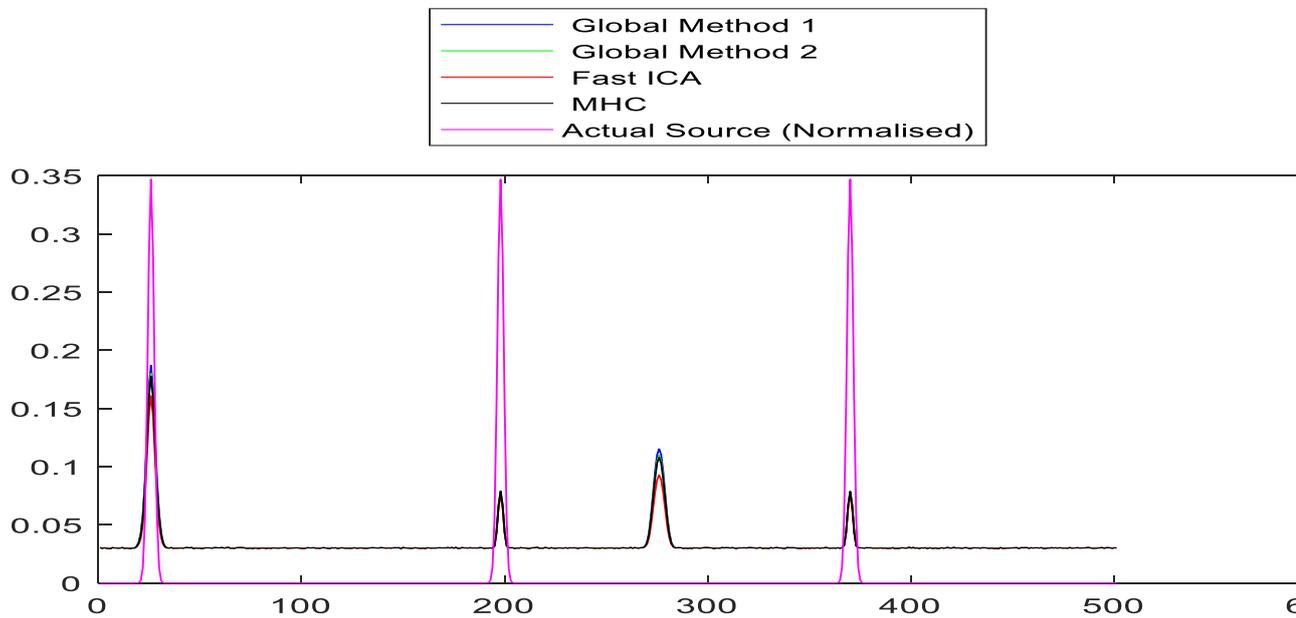

Figure 15 (Source 2) – Comparison of RMS Errors as a function of sample number for Source 2 (a) rms errors for Source 2 (b) rms errors for Source 2 along with clean signal.

For the estimation of Source 1, all three phase space methods perform almost equally well. The RMS errors become relatively large around sample numbers 198 and 370 indicating the contamination of Source 1 estimates by Source 2. The FastICA method, for this example, works best with Gaussian non-linearity and worst when using the pow3 non-linearity. When using FastICA to estimate Source 1, there is less contamination by Source 2 estimates compared to the other methods, which is indicated by lower rms errors around sample points 198 and 370 compared with the phase space methods.

For the estimation of Source2, the FastICA method has the best performance. There is no significant differences between the performances of the different phase space methods.

## 4.3     Non-Sparse Sources

Sound signals, for example speech and music, can in many cases be considered as approximately sparse, so it would be of interest to see if the proposed method can separate out the individual sources from mixtures of sound signals. The sources will, in general, be overlapping and correlated. The set of sources that we will use are taken from [29]. Further details concerning these data can be found in References [30] and [31].   The pure sources are shown in Figure 16 below.  It should be noted that the amplitude of Source 2 is significantly smaller than for Sources 1 and 3. The MHC, Global 1 and Global 2 algorithms, which rely on identifying local sparse segments of the sources are expected to do worse than FastICA which processes the data globally.

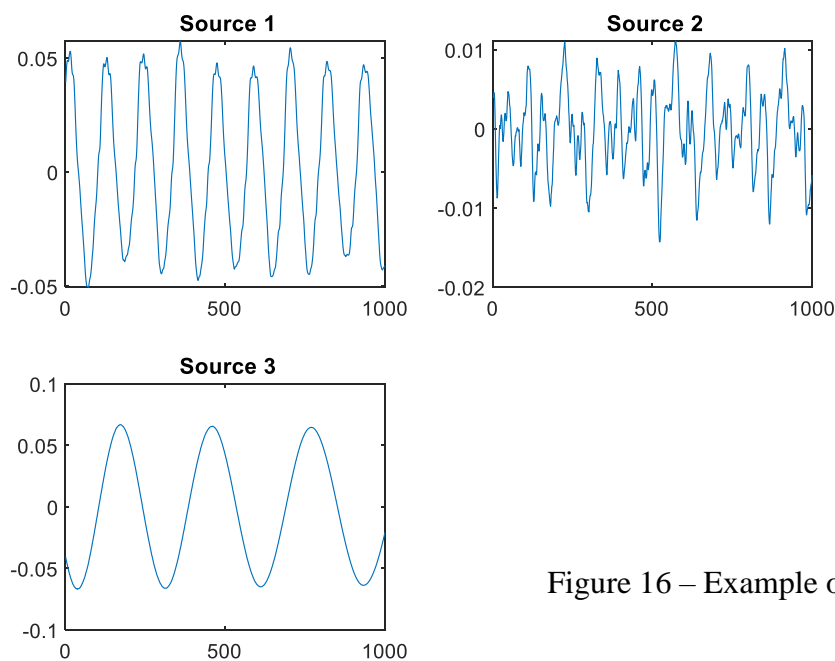

Figure 16 – Example of non-sparse sources

For the FastICA, Gaussian non-linearity is found to give the best results.

First, we are going to compare the performances of the four algorithms under test when applied to the clean data mixtures, that is where no noise has been added to each mixed data input. As there are no random elements in the MHC, Global 1 and Global 2 algorithms, the performance on the clean data can be assessed without using Monte Carlo simulations. However, there is a random element in the FastICA algorithm, which is the initialisation of the weights; and so for the FastICA, 10000 Monte Carlo runs are carried out.

The results for the rms errors as a function of time for each source are shown in Figure 17 for the four algorithms. The values of $v^{th}$ that give the best performances for the Global 1, Global 2 and MHC are found to be 0.5, 0.8 and 0.7 respectively.

**Source 1(a)**

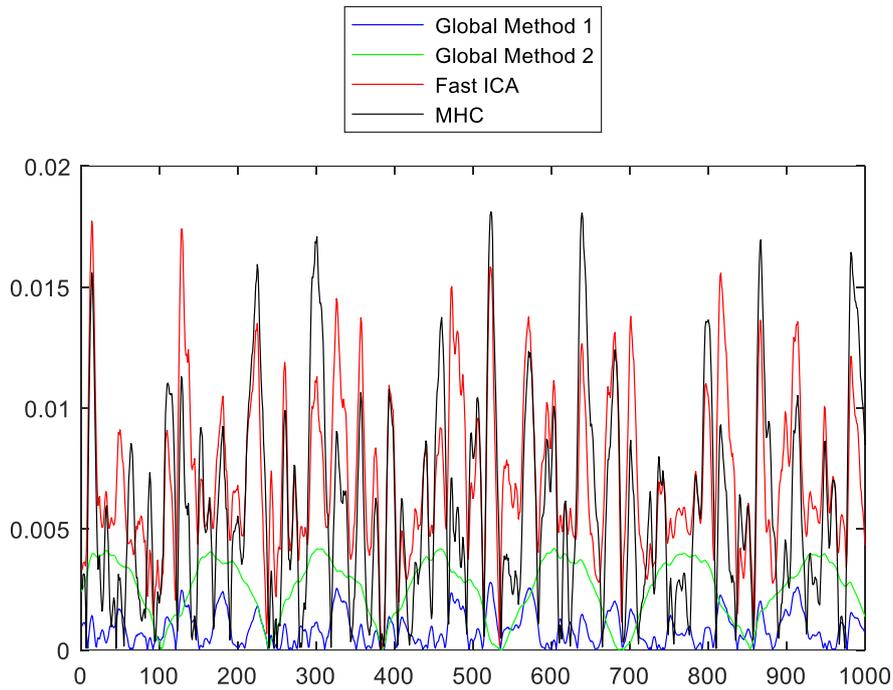

**Source 1(b)**

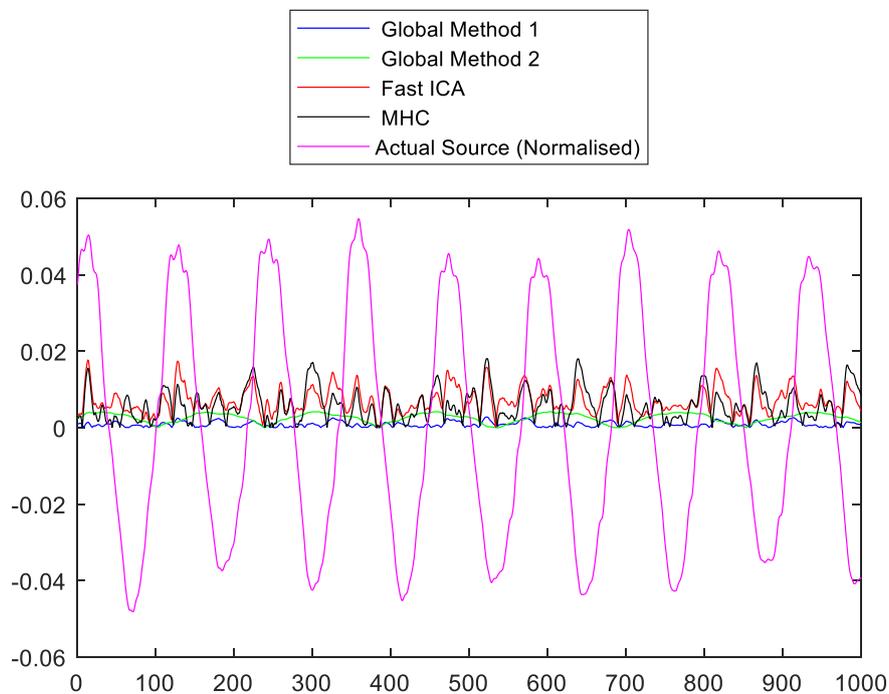

Figure 17 (Source 1) – Comparison of RMS Errors as a function of sample number for Source 1 (a) rms errors for Source 1 (b) rms errors for Source 1 along with clean signal.

**Source 2(a)**

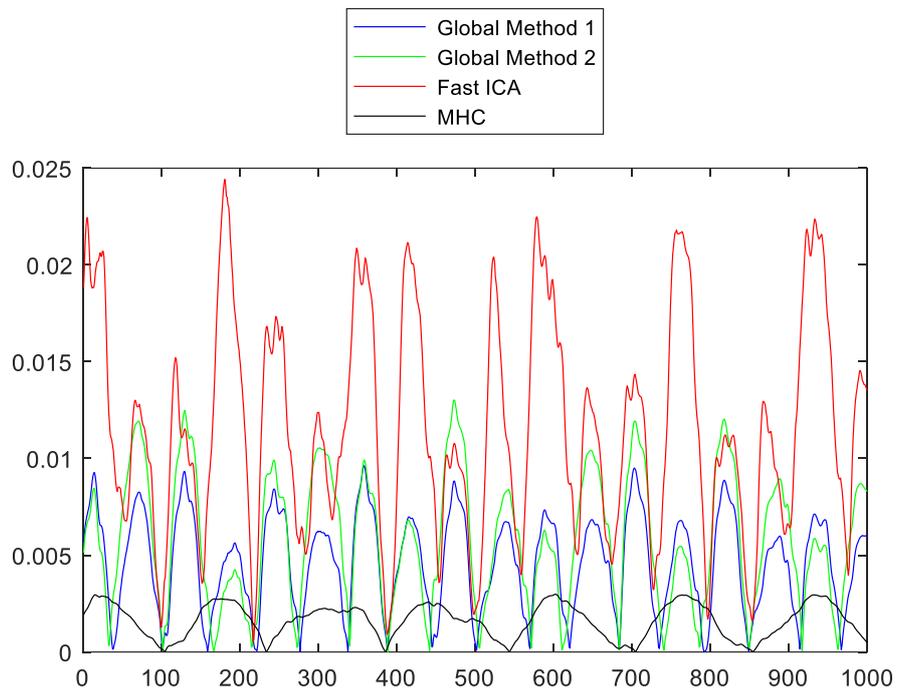

**Source 2(b)**

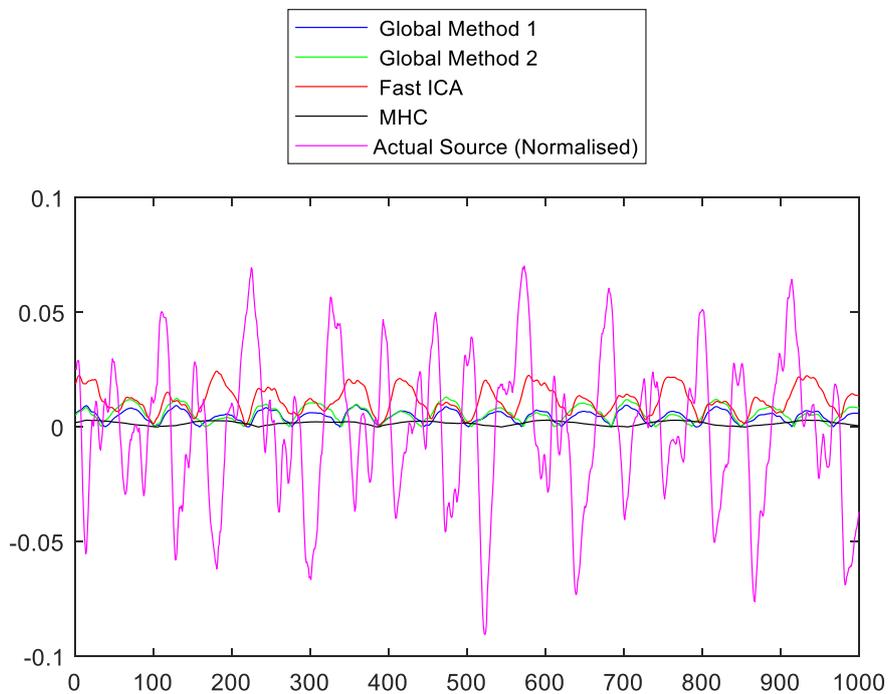

Figure 17 (Source 2) – Comparison of RMS Errors as a function of sample number for Source 2 (a) rms errors for Source 2 (b) rms errors for Source 2 along with clean signal.

**Source 3(a)**

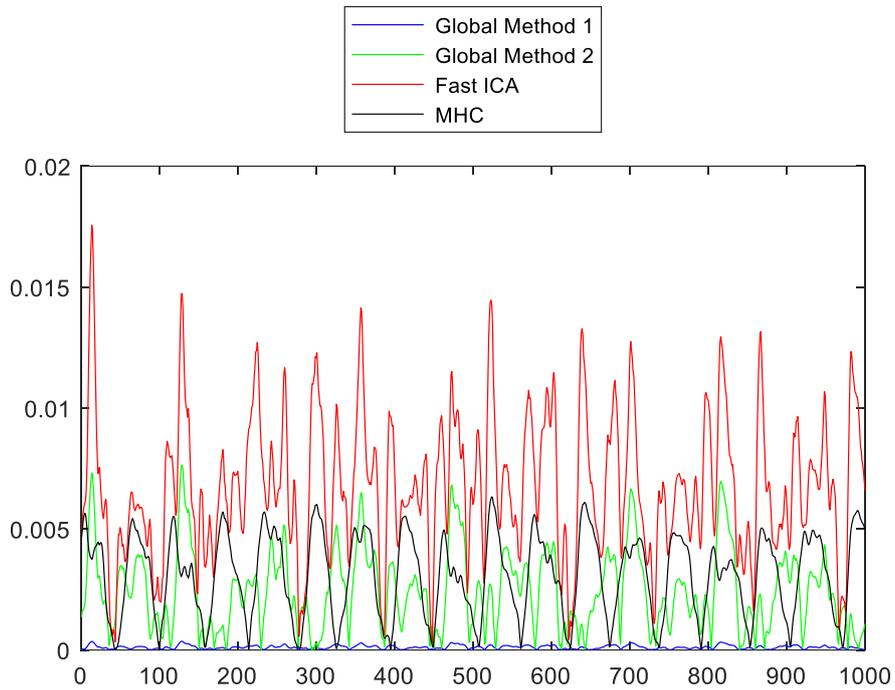

**Source 3(b)**

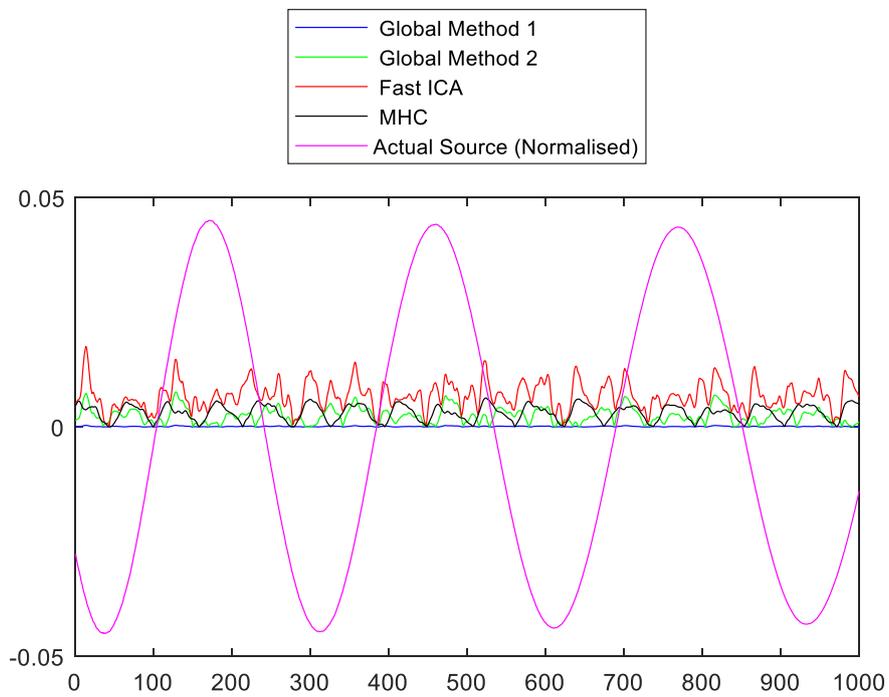

Figure 17 (Source 3) – Comparison of RMS Errors as a function of sample number for Source 3 (a) rms errors for Source 3 (b) rms errors for Source 3 along with clean signal.

There are variable performances for the four algorithms. For Sources 1 and 3, the Global 1 has the best performance. However, for Source 2, the MHC method has the best performance, which is unexpected as it is one of the methods designed specifically for mixtures of sparse sources only. The MHC method chooses headings around sample point 990; around this point, the magnitudes of the rate of changes for Sources 1 and 3 are coincidentally very similar, and the MHC hence chooses this point to compute the principal heading vector.

The methods MHC, Global 1 and Global 2 detect segments where one source is dominant, but errors occur because there are still contributions from other sources. However, Global 1 is performing an average over several headings and the errors, fortuitously, cancels out.

Overall, for these simulations, the Global 2 method has a worse performance than Global 1for estimating all three sources. Global 2 clearly does not work as well as Global 1 for these type of signal mixtures as there is no averaging of headings to cancel out errors as for Global 1. Global 1 works better because of averaging over headings, controlled by the cluster threshold (46). It is interesting to note rms errors for Global 1 tend to be lower than for FastICA, although this observation is for this set of simulations only.

It should be noted that for the estimation of a particular source, there is contamination of this estimate from other sources, even for FastICA, as the sources are not perfectly uncorrelated. This can be seen from the fact that the rms errors as a function of time for a particular source resemble the other sources.

We now look at the performances of the four algorithms when noise standard deviation of 0.0005 is added to the data mixtures. The values of $v^{th}$ for MHC, Global 1 and Global 2 are kept to be the same as used for the clean signals and Gaussian non-linearity is still used for FastICA; the results are shown in Figures18:

**Source 1(a)**

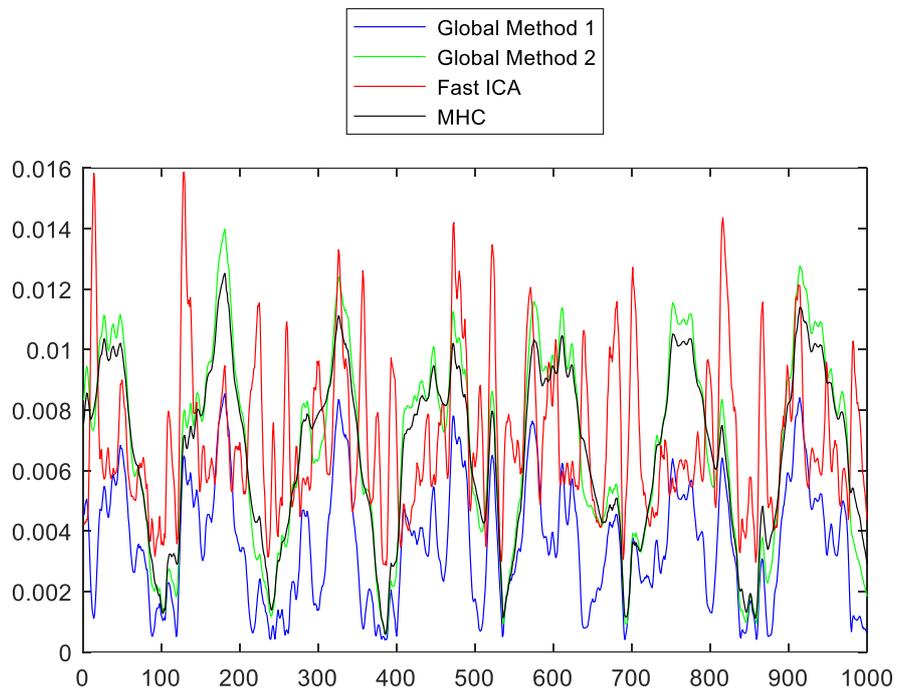

**Source 1(b)**

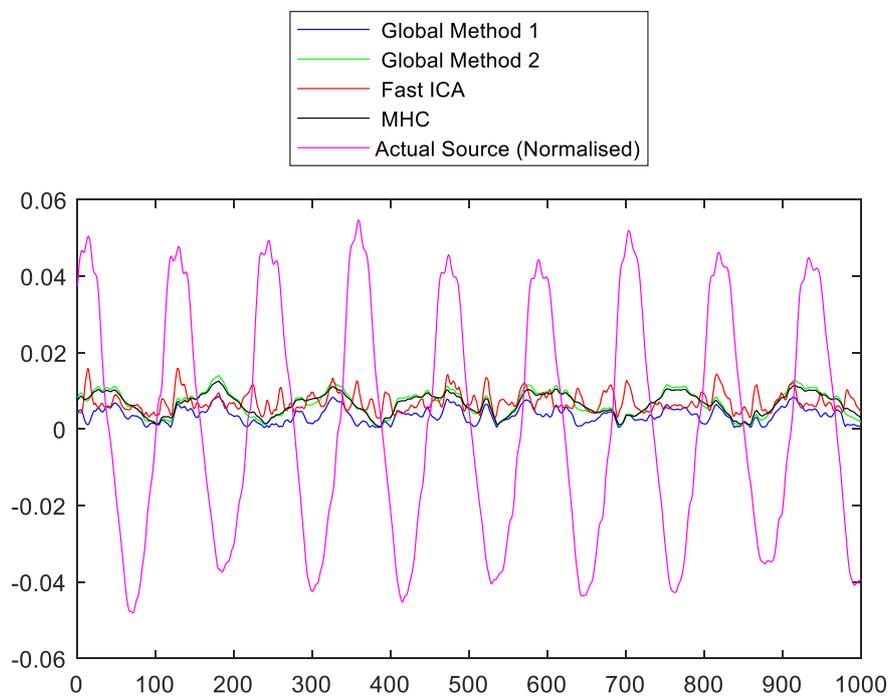

Figure 18 (Source 1) – Comparison of RMS Errors as a function of sample number for Source 1 (with noise added) (a) rms errors for Source 1 (b) rms errors with clean signal.

**Source 2(a)**

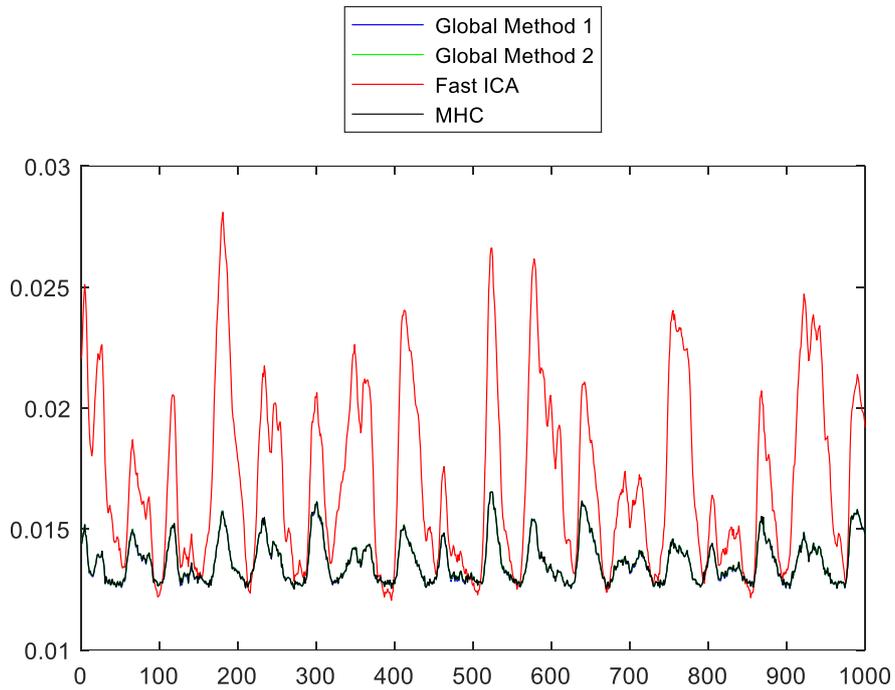

**Source 2(b)**

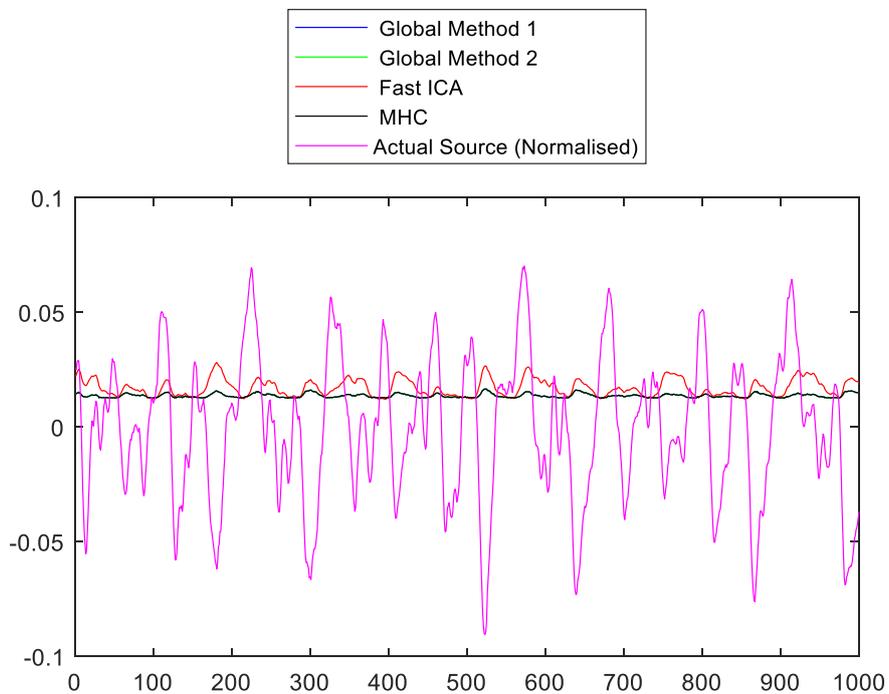

Figure 18 (Source 2) – Comparison of RMS Errors as a function of sample number for Source 2 (with noise added) (a) rms errors for Source 2 (b) rms errors with clean signal.

**Source 3(a)**

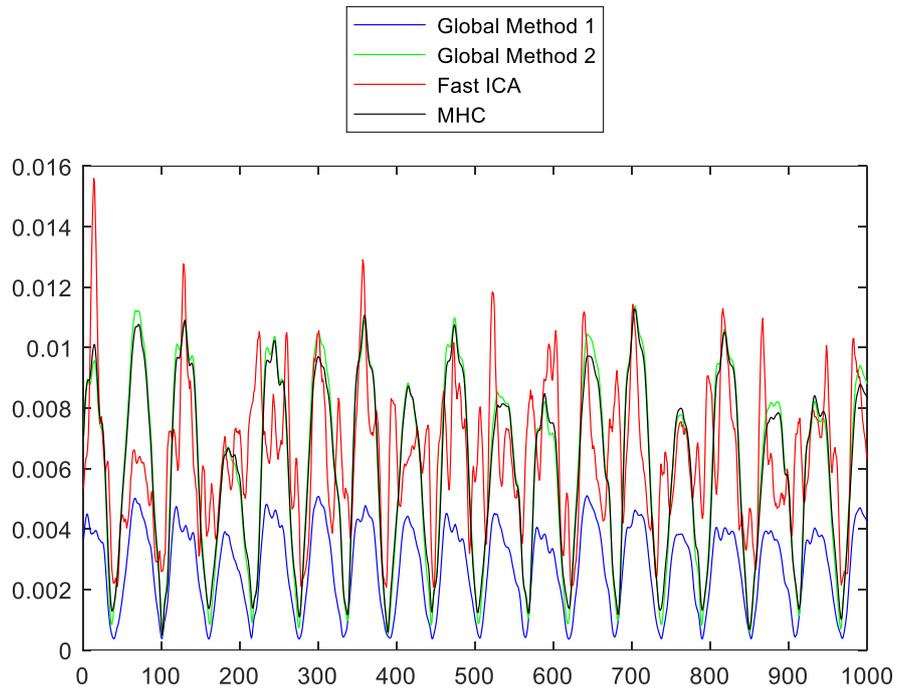

**Source 3(b)**

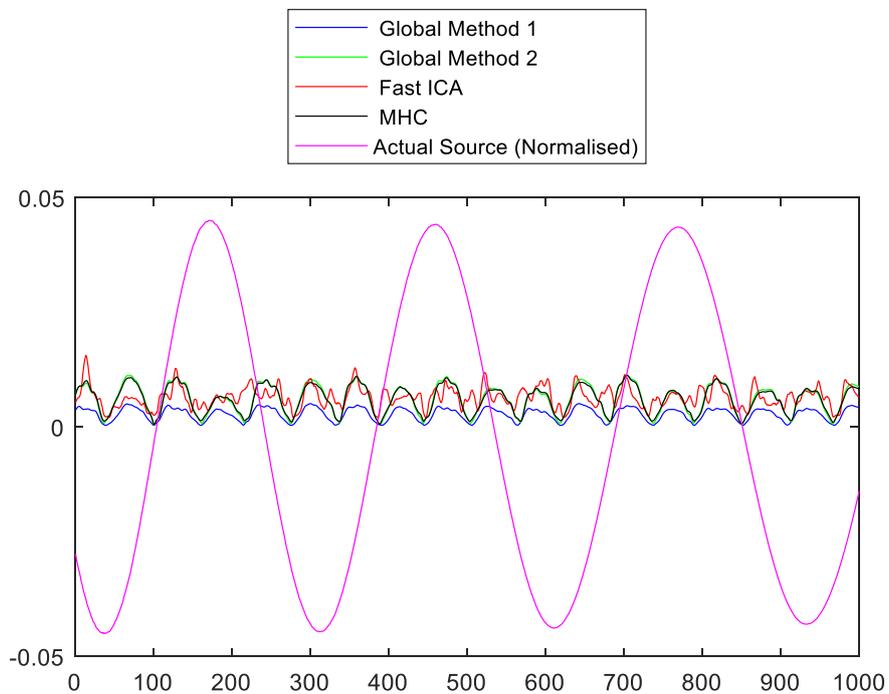

Figure 18 (Source 3) – Comparison of RMS Errors as a function of sample number for Source 3 (with noise added) (a) rms errors for Source 3 (b) rms errors with clean signal.

As for the simulations with the clean signal, the performance of the Global 1 method has overall the best performance, although for Source 1 all three phase space methods have very similar performances. Note that, as with the clean data, for each source estimate, there is contamination from other sources, which is as expected given the presence of noise and the fact that the sources are correlated.

## 4.4    8-Lead Thoracic and Abdominal ECG Data from Expectant Mother

To compare the performances of the FastICA and Global methods, data taken from the Daisy Database [32] will be analysed. This data consists of ECG signals taken from an expectant mother. The data consists of 8 leads, 1 to 5 being abdominal and 6 to 8 thoracic.

The first 1000 samples of the data are chosen; there is some uncertainty about the sampling frequency, also pointed out in [33], but it is probably 250 Hz.

Data from a typical abdominal lead is shown in Figure 19.

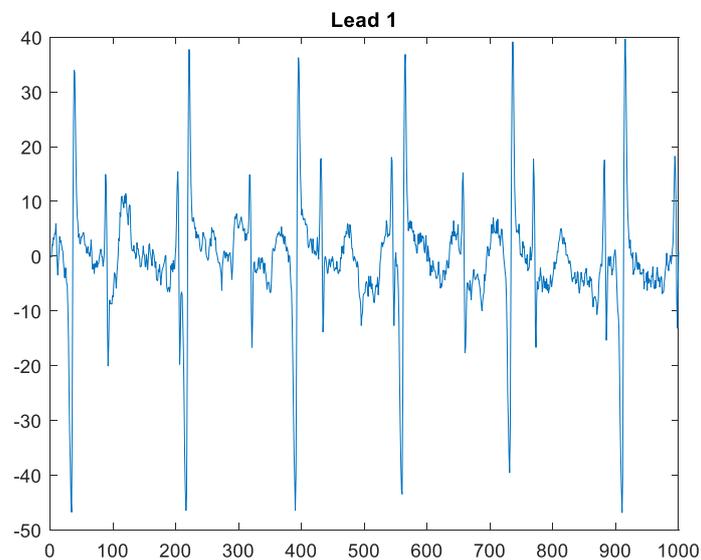

Figure 19 – Lead 1 Abdominal Signal taken from [32]

The Global 2 method has been applied to this data for various values of $v^{th}$ in (46). The best fetal extraction, from a qualitative point of view, is when $v^{th} = 0.99$. In this case, the estimate that is closest to a fetal signal is shown in Figure 20.

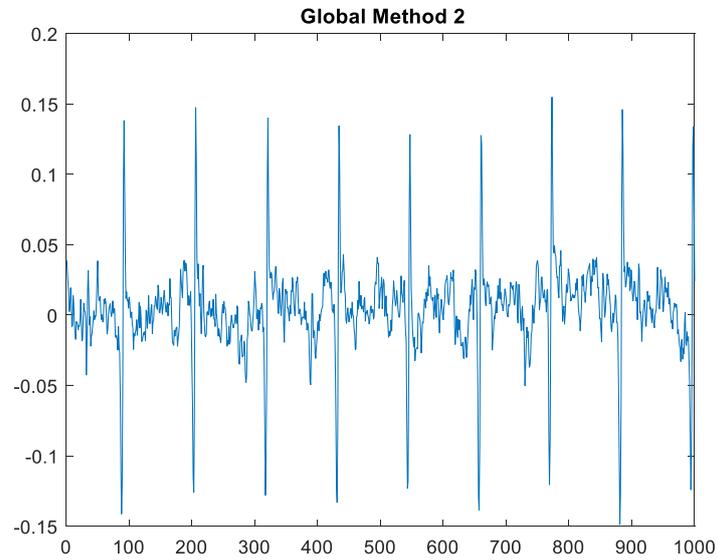

Figure 20 – Extracted fetal signal using Global 2 $v^{th} = 0.99$.

This is very close to the result obtained using the Fast ICA, shown in Figure 21.

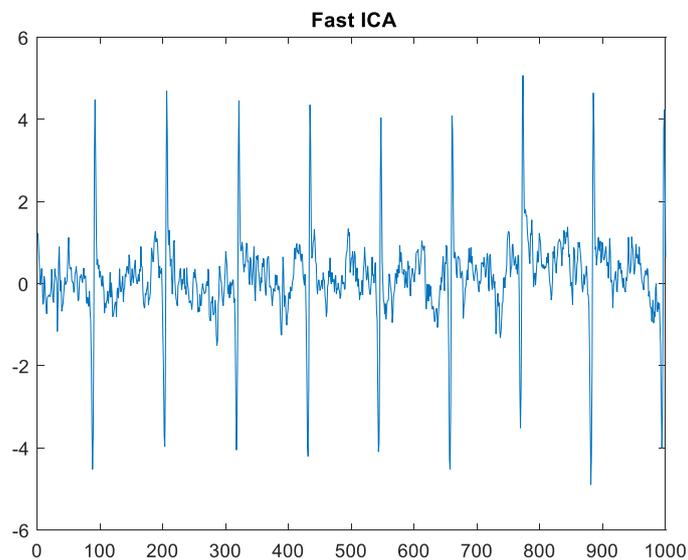

Figure 21 – Extracted fetal signal using FastICA

The quality of the estimate of the fetal component when using Global 2 is sensitive to the choice of $v^{th}$; for example, if $v^{th}$ is reduced to $0.9$, then the following fetal output is obtained:

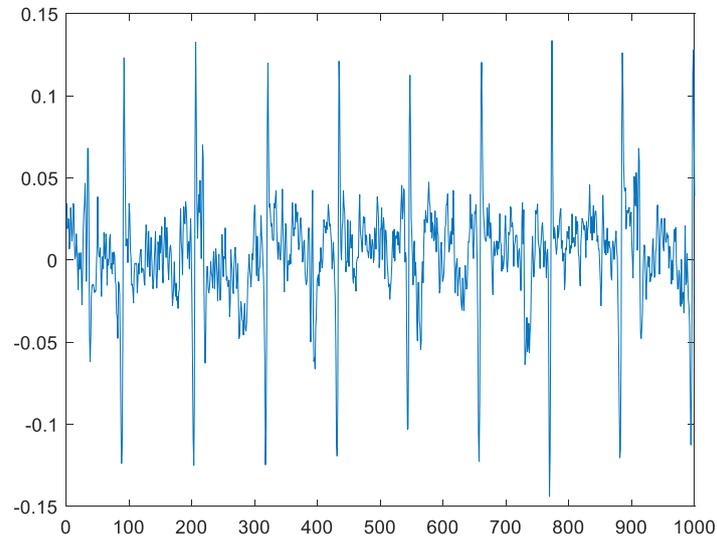

Figure 22 -Extracted fetal signal using Global 2 $v^{th} = 0.9$.

Comparing Figures (20) and (22), the estimated fetal signal is noisier when $v^{th} = 0.9$ than it is when $v^{th} = 0.99$ as velocity vectors with smaller magnitudes are more affected by noise.

## 5   Summary of Results

In this paper, we are comparing two types of algorithm for sparse component analysis: (1) algorithms that are designed for blind source separation for general sources: FastICA comes under this category (2) algorithms designed specifically for semi-blind source separation of sparse sources: MHC, Global 1 and Global 2 come under this category.

Looking at the simulations in Section 4.1, the MHC, Global 1 and Global 2 methods perform better than FastICA for purely sparse sources. The main reason for this is that FastICA subtracts the mean of the data as a preprocessing step, which can introduce correlations between the sources that were hitherto uncorrelated. The same observation can be made for partially overlapping sources in Section 4.2

In the case where there are mixtures of sources that are not sparse, Section 4.3, then, of the three phase space methods that are used, Global 1 performs the best overall as this is grouping together heading vectors and the averaging reduces the effects of noise. For the analysis of the abdominal signal, Section 4.4, the new clustering method Global 2 performs as well as FastICA, but it is important to choose carefully the $v^{th}$ parameter in (46) for the former method.

## 6 Discussion

The results presented in Section 4 are representative of results obtained with other data mixtures, where the sources are sparse, sparse & overlapping, and non-sparse.

We are particularly looking at the performance of Global 2 as a simplified alternative to Global 1, where only one input parameter is used in the former method.

The methods Global 1 and Global 2 work best where sparsity is along a particular segment or segments of the data mixtures. FastICA makes use of the whole data to maximise the independence, but this means not making use of sparse segments; this can lead to FastICA giving suboptimal results for mixtures of purely sparse sources.

However, in noise, Global 1 and Global 2 only process part of the data and try to smooth headings by Maximum Likelihood averaging (Global 1) or computing a single principal vector (Global 2) and so do not process the whole data to reduce noise. This means that these two methods can perform worse than FastICA in the presence of noise. Also, the high pass nature of differencing velocity vectors (43), increases the sensitivity of the MHC, Global1 and Global 2 methods to noise.

The FastICA method has several input parameters, but it has been found that the performance of this method is insensitive to changes in these parameters. The exception here is the non-linearity, where the optimum non-linearity can depend on the nature of the sources. The MHC and Global 2 have one input parameter, $v^{th}$ (46), which is a way of thresholding the magnitude of the velocity vectors to increase robustness of the method to noise. For Global 1, there is an additional parameter (61) for clustering the headings, although a value around $\alpha=1$ has been found to work across a variety of data mixtures.

## 7 Conclusions

Four methods have been compared to estimate various types of sources from data mixtures.

FastICA [6,7], MHC [17] and Global 1 [25] are existing methods. Global 2 has been derived in this paper as an alternative to Global 1, simplified as only one input parameter needs to be chosen.

It is unclear whether any of these methods can be considered the "best". Each method has its advantages and disadvantages. Two of the methods, Global 1 and Global 2, involve clustering of heading vectors in the phase plot of the data; both methods have been found to have a comparable performance across data, where sources are sparse or partially sparse. For the non-sparse simulations (Section 4.3) Global 1 works better than Global 2 because of the averaging over headings of the former method.

The methods of clustering used for Global 1 and Global 2 are different to the conventional methods used. These clustering methods are used in the deflation approach where the sources are estimated one at a time prior to subtraction from the data so just one principal heading is required at each iteration.

Given the sensitivity to input parameters of MHC, Global 1 and Global 2 methods, it is recommended that these are applied in offline applications, where the input parameters (46) and (61) can be varied to achieve optimum performance.